\documentclass[
aps,
prc,
superscriptaddress,
floatfix,
nofootinbib,
twocolumn,
eqsecnum
]{revtex4-2}

\usepackage{amsmath,amssymb,amsfonts,physics,graphicx,float}
\graphicspath{{pic/}}
\usepackage[setpagesize=false]{hyperref}
\allowdisplaybreaks[4]
\usepackage{graphicx}
\usepackage[labelformat=simple]{subcaption} % 図の下のがカッコなし。
\usepackage{ulem}
\usepackage{color}
\usepackage{xcolor}
\usepackage{slashed}
\definecolor{ar}{rgb}{1.0, 0.01, 0.24}
\definecolor{al}{rgb}{0.82, 0.1, 0.26}
\definecolor{ev}{rgb}{0.56, 0.0, 1.0}

\begin{document}

\title{
Study of pion condensate in nuclear matter based on a Parity Doublet Model}

%% Write your mail %%%
\author{Xiang Liu}
\email{liu.xiang.f6@s.mail.nagoya-u.ac.jp}
\affiliation{Department of Physics, Nagoya University, Nagoya 464-8602, Japan}

\author{Bikai Gao}
\email{bikai@rcnp.osaka-u.ac.jp}
\affiliation{Research Center for Nuclear Physics (RCNP), Osaka University, Osaka 567-0047, Japan}

\author{Masayasu Harada}
\email{harada@hken.phys.nagoya-u.ac.jp}
\affiliation{Kobayashi-Maskawa Institute for the Origin of Particles and the Universe, Nagoya University, Nagoya, 464-8602, Japan}
\affiliation{Department of Physics, Nagoya University, Nagoya 464-8602, Japan}
\affiliation{Advanced Science Research Center, Japan Atomic Energy Agency, Tokai 319-1195, Japan}

\date{\today}

% I think you can write the abstrac & conclusion at last, write the main part first
\begin{abstract}
In this work, we investigate the possibility of pion condensate in isospin-asymmetric nuclear matter within a parity doublet model at zero temperature, treating the baryon and isospin chemical potentials as independent variables. We include the $\sigma$, $\pi$, $\omega$, and $\rho$ mesons and determine their mean fields from the gap equations. We introduce a novel subtraction scheme to calculate the thermodynamic potential in the no-sea approximation, in which the effective isospin chemical potential is retained in the subtraction term to remove the divergence. We find that the presence of baryons suppresses pion condensate, and that transition to the pion condensation phase is of second order.
We further find that the phase boundary depends systematically on the chiral-invariant mass $m_0$: a smaller $m_0$ requires a larger critical isospin chemical potential for pion condensate. The presence of pion condensate also shifts the nuclear Liquid-Gas phase boundary toward larger baryon chemical potentials. Finally we study the existence of pion condensation phase in neutron star matter and find that pion condensate is unlikely to occur.
\end{abstract}
\maketitle

%\tableofcontents

%%%%%%%%%%%%%%%%%%%%%%%%%%%%%%%%%%%%%%%%%%%%%%%%%%%%%%%%%%%%%%%%
\section{Introduction}
The origin of nucleon mass remains one of the most challenging frontiers in quantum chromodynamics (QCD)---the fundamental theory to describe strong interactions---and nuclear physics. A nucleon weighs around 1 GeV but three quarks that make this nucleon only weigh around 10 MeV. In the QCD vacuum, the $SU(2)_L \times SU(2)_R$ chiral symmetry spontaneously breaks down to $SU(2)_V$ by the quark-antiquark condensate, which generates nucleon mass. However, Lattice QCD simulations~\cite{Aarts:2015mma,Aarts:2017rrl,Aarts:2018glk} and QCD sum rule studies~\cite{PhysRevD.103.L051501, PhysRevD.105.014014, sym15040799} suggest that spontaneous chiral symmetry breaking does not account for the entire nucleon mass. In particular, a nonzero component of the nucleon mass may persist even after chiral symmetry is restored. This idea is adopted by the parity doublet model (PDM) \cite{Detar:1988kn, Jido:2001nt}, in which the nucleon mass is divided into a chiral variant contribution generated by spontaneous chiral symmetry breaking and a chiral invariant contribution that remains even in the chiral symmetric phase.

To explore chiral invariant mass, strongly interacting matter  under extreme conditions, such as high temperature or high density,  provides a particularly useful setting. For instance, in high density regime, chiral symmetry is partially restored and chiral condensate is reduced, while the chiral invariant mass remains. In the past few years, there are many studies aiming on how to constrain chiral invariant mass based on neutron star observational data \cite{Yamazaki:2019tuo,Minamikawa:2020jfj,Minamikawa:2021fln, Gao:2022klm, Kong:2023nue, Minamikawa:2023eky, Gao:2024chh,Gao:2025nkg}.
In particular, Refs.~\cite{Minamikawa:2020jfj, Minamikawa:2021fln, Gao:2022klm, Kong:2023nue, Minamikawa:2023eky, Gao:2024chh,Gao:2025nkg} 
use a PDM-NJL crossover model, in which neutron star equation of state is constructed
by combining low density hadronic matter made from the parity doublet model and high density quark matter from Nambu-Jona-Lassino type quark model. Comparing the resultant mass-radius relation with neutron star observational data, 
it is found that the chiral invariant mass is more than half of nucleon mass.

However, for simplicity, pion condensation has not been included in previous studies 
based on PDMs. Pion is a meson with a relatively small mass $m_\pi \approx 140\,\mathrm{MeV}$, and carries isospin number. Its chemical potential $\mu_\pi$ is equal to the isospin chemical potential $\mu_I$, $\mu_\pi = \mu_I$. Since neutron star matter is highly neutron rich, 
$\mu_I$ can be large enough as 
$\mu_I > m_\pi$, so that a Bose-Einstein condensate  may form. In other words, pion condensate is possible to occur in neutron star matter. Moreover, previous and recent studies indicate the possible existence of pion condensate based on different models \cite{Sawyer:1972cq, Scalapino:1972fu, Migdal:1973zm, Baym:1974vzp, BROWN19761, Kojo:2024sca, Voskresensky:2022gts, Fore:2023gwv, 94zt-q2dw, Motta:2026pby}.

In this work, we investigate whether pion condensation occurs within the PDM 
constructed in Ref.~\cite{Motohiro:2015taa} and characterize its properties by studying isospin-asymmetric nuclear matter at zero temperature, with the isospin chemical potential $\mu_I$ and baryon chemical potential $\mu_B$ treated as independent parameters. 
For this purpose, we introduce four mesons $\sigma,\pi,\omega$ and $\rho$ to include the attractive and repulsive force. In particular, vector mesons $\omega$ and $\rho$ are introduced based on the 
hidden local symmetry (HLS) \cite{BANDO1988217,HARADA20031} to keep the Lagrangian chiral invariant. Regarding the determination of the model parameters, we assume that, at the saturation density, there is no pion condensate and fit the model parameters to the nuclear saturation properties.

After the model parameters are fixed as in the previous analyses~\cite{Minamikawa:2020jfj,Minamikawa:2021fln,Gao:2022klm,Kong:2023nue,Minamikawa:2023eky,Gao:2024chh,Gao:2025nkg},
we calculate the thermodynamic 
potential in the no-sea approximation. Here, we propose a new subtraction scheme in which we keep the effective isospin chemical potential $\mu_I^*$ in the subtraction term to cancel the divergence. Then, we study the phase structure of asymmetric nuclear matter and show the existence of  pion condensation phase.

This paper is structured as below. In Sec.~\ref{sec:formulation} we briefly introduce the parity doublet model and write down the Lagrangian. 
In Sec.~\ref{sec:matter} we construct the nuclear matter with a non-zero pion field and employ a new subtraction scheme. In Sec.~\ref{sec:phase} we investigate the phase structure of nuclear matter with pion condensate and present the phase boundary between the pion condensation phase and the normal phase as well as the phase boundary of 
the Liquid-Gas phase transition.
In Sce.~\ref{sec:NS} we discuss how pion condensate affect the neutron star matter in parity doublet model. Finally, in 
Sec.~\ref{sec:summary} we give a summary and discussions.

\section{Formulation}
\label{sec:formulation}

%First introduce the Lagrangian in the unitary gauge of HLS, then stress how to take mean field keep consistent with the chiral symmetry 
%when introduce one-forms, explain the existence of pi mean field which is different from before

In this section we briefly review the PDM constructed in Ref.~\cite{Motohiro:2015taa}, and determine the model parameters.

\subsection{Lagrangian and mean field approach}

First of all, we introduce two kinds of nucleons as parity partners to construct the Lagrangian, one is $N(939)$ (normal nucleon with $J^P=\frac{1}{2}^+$) and the other is $N(1535)$ (excited state with $J^P=\frac{1}{2}^-$).
Since different nucleons can belong to different representations, in this work we adopt the so-called mirror assignment \cite{Detar:1988kn, Jido:2001nt}.
The transformation properties of two fields under the chiral SU(2)$_L\times$SU(2)$_R$ symmetry are given by
\begin{align}
\psi_{1L} \to g_L \psi_{1L} \ , \quad
\psi_{1R} \to g_R \psi_{1R} \ , \notag\\
\psi_{2L} \to g_R \psi_{2L} \ , \quad
\psi_{2R} \to g_L \psi_{2R} \ ,
\end{align}
where $g_{L,R} \in \mbox{SU(2)}_{L,R}$.
%\begin{equation}
%\begin{aligned}
%    \psi_{1{\mathrm{L}}} \sim(\frac{1}{2},0), \quad & \psi_{1{\mathrm{R}}} \sim(0,\frac{1}{2}),  \\\psi_{2{\mathrm{L}}}\sim(0,\frac{1}{2}), \quad & \psi_{2{\mathrm{R}}} \sim(\frac{1}{2},0). \label{eq:mirror}
%\end{aligned}
%\end{equation}
This assignment allows the existence of the chiral invariant mass.

The kinetic and mass terms for these two fields are given by~\cite{Detar:1988kn,Jido:2001nt}
\begin{align}
\mathcal{L}_{N}= & \bar{\psi}_{1} i \slashed{D} \psi_{1}+\bar{\psi}_{2} i \slashed{D} \psi_{2}\\
&-g_{1}\left(\bar{\psi}_{1L} M \psi_{1R}+\bar{\psi}_{1R} M^{\dagger} \psi_{1L}\right) \notag \\
&-g_{2}\left(\bar{\psi}_{2L} M^{\dagger} \psi_{2R}+\bar{\psi}_{2R} M \psi_{2L}\right) \notag \\
& -m_{0}\left(\bar{\psi}_{1L} \psi_{2R}-\bar{\psi}_{1R} \psi_{2L}-\bar{\psi}_{2L} \psi_{1R}+\bar{\psi}_{2R} \psi_{1L}\right) \ .
\label{eq:N kin}
\end{align}
where
$M$ is the $2\times2$ matrix field for scalar and pseudscalar mesons, which transforms as
\begin{equation}
M \ \to \ g_L \, M \, g_R^\dag \ .
\end{equation}
In Eq.(\ref{eq:N kin}) 
%the above expression, 
the covariant derivatives acting on the nucleon fields are defined as
\begin{align}
D_\mu \psi_{1L} & = \left( \partial_\mu - i {\mathcal L}_\mu - i {\mathcal B}_\mu \right)\psi_{1L} \ , \notag\\
D_\mu \psi_{1R} & = \left( \partial_\mu - i {\mathcal R}_\mu - i {\mathcal B}_\mu\right) \psi_{1R} \ , \notag\\
D_\mu \psi_{2L} & = \left( \partial_\mu - i {\mathcal R}_\mu - i {\mathcal B}_\mu\right)\psi_{2L} \ , \notag\\
D_\mu \psi_{2R} & = \left( \partial_\mu - i {\mathcal L}_\mu - i {\mathcal B}_\mu\right) \psi_{2R} \ , 
\end{align}
where
${\mathcal L}_\mu$ (${\mathcal R}_\mu$) is the external gauge field corresponding to SU(2)$_{L}$ (SU(2)$_R$) chiral symmetry, and ${\mathcal B}_\mu$ is the external gauge field coresponding to the baryon number U(1) symmetry.

In Ref.~\cite{Motohiro:2015taa}, 
the vector mesons $\rho$ and $\omega$ are introduced based on the hidden local symmetry (HLS)~\cite{BANDO1988217,HARADA20031} by decomposing the matrix field $M$ as
\begin{equation}
M=\xi_{L}^{\dagger} \sigma \xi_{R}\ ,
\label{M decomp}
\end{equation}
where $\xi_{L}$ and $\xi_R$ transform as
\begin{align}
\xi_L \ \to \ h_\rho h_\omega \xi_L g_L^\dag \ , \notag\\
\xi_R \ \to \ h_\rho h_\omega \xi_R g_R^\dag \ ,
\end{align}
with $h_\rho$ and $h_\omega$ being the elements of SU(2)$_{\rm HLS}$ and U(1)$_{\rm HLS}$, respectively.
In the unitary gauge of the HLS, 
the variables $\xi_L$ and $\xi_R$ are 
written in terms of the pion fields $\pi_a$ ($a= 1,2,3$) as
\begin{equation}
\xi_{L}=e^{-i \sum_{a=1}^{3} \pi_{a} T^{a}}, \quad \xi_{R}=e^{i \sum_{a=1}^{3}\pi_{a} T^{a}} \ ,
\label{eq:non-linear pi}
\end{equation}
where the matrix $T^a=\tau_a/2$ with $\tau_a$ being the Pauli Matrix.
In Eq.~(\ref{M decomp}) $\sigma$ is the non-matrix field expressing the iso-singlet scalar meson, which is invariant under both the chiral symmetry and HLS. 

The relevant Maurer-Cartan one-forms 
are given by
\begin{align}
\alpha_{\mu}^{\|} & =\frac{1}{2 i}\left[\left(D_{\mu} \xi_{R}\right) \xi_{R}^{\dagger}+\left(D_{\mu} \xi_{L}\right) \xi_{L}^{\dagger}\right], \notag \\
\alpha_{\mu}^{\perp} & =\frac{1}{2 i}\left[\left(D_{\mu} \xi_{R}\right) \xi_{R}^{\dagger}-\left(D_{\mu} \xi_{L}\right) \xi_{L}^{\dagger}\right],
\end{align}
where the covariant derivatives are 
\begin{align}
D_{\mu} \xi_{L} & =\left(\partial_{\mu}-i g_{\rho} \sum_{a=1}^{3}\rho_{\mu}^{a} T^{a} -\frac{1}{2} i g_{\omega} \omega_{\mu} \right) \xi_{L}+i \xi_{L} \mathcal{L}_{\mu}, \notag \\
D_{\mu} \xi_{R} & =\left(\partial_{\mu}-i g_{\rho} \sum_{a=1}^{3}\rho_{\mu}^{a} T^{a} -\frac{1}{2} i g_{\omega} \omega_{\mu} \right) \xi_{R}+i \xi_{R} \mathcal{R}_{\mu}
\end{align}
with $\rho_\mu$ and $\omega_\mu$ being HLS gauge fields for SU(2)$_{\rm HLS}$ and U(1)$_{\rm HLS}$, and $g_\rho$ and $g_\omega$ their gauge couplings, respectively.
The Lagrangian for vector mesons is given by
\begin{align}
\mathcal{L}_{\text {vector }}= & \, -\frac{1}{4} \omega_{\mu \nu} \omega^{\mu \nu}
- \frac{1}{2} \operatorname{tr}\left(\rho_{\mu \nu} \rho^{\mu \nu}\right)
\notag\\
& \, {} +\frac{m_{\omega}^{2}}{2 g_{\omega}^{2}} \operatorname{tr}\left(\hat{\alpha}_{\mu}^{\|}\right) \operatorname{tr}\left(\hat{\alpha}_{\|}^{\mu}\right) \notag\\
&\, {}+ \frac{m_{\rho}^{2}}{g_{\rho}^{2}}\left(\operatorname{tr}\left(\hat{\alpha}_{\mu}^{\|} \hat{\alpha}_{\|}^{\mu}\right)-\frac{1}{2} \operatorname{tr}\left(\hat{\alpha}_{\mu}^{\|}\right) \operatorname{tr}\left(\hat{\alpha}_{\|}^{\mu}\right)\right)\ , 
\end{align}
where $m_\omega$ and $m_\rho$ are the masses of $\omega$ and $\rho$ mesons, and
$\omega_{\mu\nu}$ and $\rho_{\mu\nu}$ are the gauge field strengths of the HLS gauge bosons.
The Lagrangian for the interactions to the vector mesons is expressed as~\cite{Motohiro:2015taa}
\begin{align}
\label{eq:lagrangian_intvec}
\mathcal{L}_{NV}= &\, a_{V N N}\bigg[\bar{\psi}_{1L} \xi_{L}^{\dagger} \gamma^{\mu} \hat{\alpha}_{\mu}^{\|} \xi_{L} \psi_{1L}+\bar{\psi}_{1R} \xi_{R}^{\dagger} \gamma^{\mu} \hat{\alpha}_{\mu}^{\|} \xi_{R} \psi_{1R} \notag \\
&\qquad {}+\bar{\psi}_{2L} \xi_{R}^{\dagger} \gamma^{\mu} \hat{\alpha}_{\mu}^{\|} \xi_{R} \psi_{2L}+\bar{\psi}_{2R} \xi_{L}^{\dagger} \gamma^{\mu} \hat{\alpha}_{\mu}^{\|} \xi_{L} \psi_{2R}\bigg] \notag \\
& +a_{0 N N} \sum_{i=1,2}\bigg[\bar{\psi}_{iL} \gamma^{\mu} \operatorname{tr}\left(\hat{\alpha}_{\mu}^{\|}\right) \psi_{iL}\notag\\
& \qquad\qquad\qquad {}+\bar{\psi}_{iR} \gamma^{\mu} \operatorname{tr}\left(\hat{\alpha}_{\mu}^{\|}\right) \psi_{iR}\bigg].
\end{align}
%%%%%%%%%%%%%%%%%%

For obtaining the thermodynamic potential to construct nuclear matter, we take mean-field approximation.
As in the previous analyses
(see e.g. Ref.~\cite{Gao:2025nkg}), 
we introduce the chemical potentials for the baryon number and the isospin through the time component of the external gauge fields as
\begin{align}
{\mathcal B}_\mu & = \mu_B \, \delta_{\mu0} \ , 
\notag\\
\mathcal{R}_{\mu} & = \mathcal{L}_{\mu}=\mu_{I} T^{3} \delta_{\mu 0}.
\end{align}
We assume that
the rotational symmetry of the system is not broken, so that vector mesons $\omega$ and $\rho$ have nonzero mean field values
only for the time components. Furthermore, 
due to the isospin symmetry, we assume that the $\rho$ meson has a nonzero mean field value only for the third component.
In this study, we consider only the S-wave pion condensate as a first step. Since $\pi^0$ meson does not feel isospin chemical potential $\mu_I$ and baryon chemical potential $\mu_B$, it does not condense in the matter. 
In neutron star matter, the isospin chemical potential $\mu_I$ is negative, so that one can expect that $\pi^-$ meson condenses.
The existence of a finite isospin chemical potential 
explicitly breaks the isospin SU(2) symmetry to $U(1)_{I_3}$, which allows us to rotate $\pi^-$ condensate to the $\pi_1$ direction in the mean field approach. Now, we introduce the mean fields $\pi$, $\omega$ and $\rho$ as 
\begin{align}
%\sigma=\sigma_0, \  
& \pi_a=\pi \delta_{a1}, \ \omega_\mu=\omega \delta_{\mu0}, \notag\\
& \rho_\mu^a= \delta_{\mu0}\delta^{a3}\left( \rho + \frac{\mu_I}{g_\rho} \mu_I \right)
 \ ,
\label{eq:mean fields definition}
\end{align}
and use $\sigma$ for the mean field for $\sigma$ meson.

\iffalse
\begin{equation}
M=\xi_{L}^{\dagger} \sigma \xi_{R},
\end{equation}
where
\begin{equation}
\xi_{L}=e^{-i \pi_{1} T^{1}}, \quad \xi_{R}=e^{i \pi_{1} T^{1}}.
\end{equation}
To construct the Lagrangian, we introduce vector mesons $\omega$ and $\rho$ following Hidden Local Symmetry \cite{HARADA20031} and the relevant Cartan-Maurer one-forms are given by
\begin{align}
\alpha_{\mu}^{\|} & =\frac{1}{2 i}\left[\left(D_{\mu} \xi_{R}\right) \xi_{R}^{\dagger}+\left(D_{\mu} \xi_{L}\right) \xi_{L}^{\dagger}\right], \notag \\
\alpha_{\mu}^{\perp} & =\frac{1}{2 i}\left[\left(D_{\mu} \xi_{R}\right) \xi_{R}^{\dagger}-\left(D_{\mu} \xi_{L}\right) \xi_{L}^{\dagger}\right],
\end{align}
where the covariant derivatives are 
\begin{align}
D_{\mu} \xi_{L} & =\left(\partial_{\mu}-i g_{\rho} \rho_{0}^{3} T^{3} \delta_{\mu 0}-\frac{1}{2} i g_{\omega} \omega_{0} \delta_{\mu 0}\right) \xi_{L}+i \xi_{L} \mathcal{L}_{\mu}, \notag \\
D_{\mu} \xi_{R} & =\left(\partial_{\mu}-i g_{\rho} \rho_{0}^{3} T^{3} \delta_{\mu 0}-\frac{1}{2} i g_{\omega} \omega_{0} \delta_{\mu 0}\right) \xi_{R}+i \xi_{R} \mathcal{R}_{\mu}.
\end{align}
According to the rotational symmetry of the system, vector mesons have nonzero mean field values only for the time components. The attractive force provided by the scalar meson and the repulsive force provided by the vector mesons are balanced to create stable nuclear matter. For $\rho$ meson, due to the isospin symmetry, we assume it has a nonzero mean field values only for the third component as an iso-vector.
\fi

With mean-field values and external chemical potentials, the 
Maurer-Cartan 
one-forms are calculated as
%\begin{equation}
%\begin{aligned}
%\hat{\alpha}_{\mu}^{\|}&=-\left(g_{\rho} \rho_{0}^{3}-\mu_{I} \cos \pi_1 \right) T^{3} \delta_{\mu 0}-\frac{1}{2} g_{\omega} \omega_{0} \delta_{\mu 0},  \\
%\hat{\alpha}_{\mu}^{\perp}&=\mu_{I} T^{2} \sin \pi_1 \delta_{\mu 0}.
%\end{aligned}
%\end{equation}
\begin{align}
\hat{\alpha}_{\mu}^{\|}&=- \left( g_\rho \rho T^{3} +\frac{1}{2} g_{\omega} \omega \right) \delta_{\mu 0}\, ,  \\
\hat{\alpha}_{\mu}^{\perp}&=\mu_{I} T^{2} \sin \pi \delta_{\mu 0}\, . 
\end{align}
%\begin{align}
%\operatorname{tr}\left(\hat{\alpha}_{\mu}^{\|}\right) \operatorname{tr}\left(\hat{\alpha}_{\|}^{\mu}\right) & =\left(g_{\omega} \omega_{0}\right)^{2}, \notag \\
%\operatorname{tr}\left(\hat{\alpha}_{\mu}^{\|} \hat{\alpha}_{\|}^{\mu}\right)-\frac{1}{2} \operatorname{tr}\left(\hat{\alpha}_{\mu}^{\|}\right) \operatorname{tr}\left(\hat{\alpha}_{\|}^{\mu}\right) & = \frac{1}{2}\left(g_{\rho} \rho_{0}^{3}-\mu_{I} \cos \pi_1 \right)^{2}. \notag
%\end{align}

Then the vector meson Lagrangian is reduced to
\begin{equation}
- \mathcal{L}_{\text {vector }}
\ \to \ V_{\rm vector} =  - \frac{1}{2} m_{\omega}^{2} \omega^{2} - \frac{1}{2} m_{\rho}^{2} \rho^2.
\end{equation}

In addition, the kinetic term of scalar mesons and the symmetry breaking term are expressed as 
\begin{align}
\mathcal{L}_{M}^{\rm kin} &= 
\frac{1}{4} \mbox{tr} \left[ D_\mu M^\dag D^\mu M \right]
%\sigma^{2} \operatorname{tr}\left(\alpha_{\mu}^{\perp} \alpha_{\perp}^{\mu}\right)
\ \to \ 
%= 
%\frac{1}{2}\sigma^2\partial_\mu \pi \partial^\mu \pi+
\frac{1}{2} \sigma^{2} \mu_{I}^{2} \sin ^{2} \pi, \notag \\
V_{\rm SB} &=  -\frac{1}{4} m_{\pi}^{2} f_{\pi} \operatorname{tr}\left[M+M^{\dagger}\right]
\ \to \ -\sigma f_{\pi} m_{\pi}^{2} \cos \pi.
\label{eq:VSB}
\end{align}
In the Lagrangian Eq.~(\ref{eq:lagrangian_intvec}), nucleon fields $\psi$ often appears together with $\xi$.
For simplicity of the calculation, we perform the chiral transformation as below to absorb $\xi$:
\begin{align}
N_{1 L}=\xi_{L} \psi_{1 L}, & \quad N_{1 R}=\xi_{R} \psi_{1 R}, \notag \\
N_{2L}=\xi_{R} \psi_{2L}, & \quad N_{2 R}=\xi_{L} \psi_{2R} .\notag
\end{align}
The Lagrangian is finally written as
\begin{equation}
\begin{aligned}
\mathcal{L} %& =\mathcal{L}_{k i n}+\mathcal{L}_{I}  \\ 
& =\bar{N}_{1}\left(i \slashed{\partial}-g_{1} \sigma+\left(\mu_{B}-g_{\omega N N} \omega\right) \gamma^{0} \right.\\
&\quad \left.+\left(\mu_{I} \cos \pi-g_{\rho N N} \rho\right) T^{3} \gamma^{0}+\left(\mu_{I} \sin \pi\right) T^{2}\gamma^{0}\gamma^{5}\right) N_{1}  \\
&\quad +\bar{N}_{2}\left(i \slashed{\partial}-g_{2} \sigma+\left(\mu_{B}-g_{\omega N N} \omega \right) \gamma^{0}\right.\\
& \quad\left.+\left(\mu_{I} \cos \pi-g_{\rho N N} \rho \right) T^{3} \gamma^{0}-\left(\mu_{I} \sin \pi\right) T^{2} \gamma^{0} \gamma^{5}\right) N_{2}  \\
&\quad -m_{0}\left(\bar{N}_{1} \gamma^{5} N_{2}-\bar{N}_{2} \gamma^{5} N_{1}\right)-V_{{\rm meson} },
\label{eq:finallagrangian}
\end{aligned}
\end{equation}
with the meson part
\begin{align}
&V_{{\rm meson}}=V_{{\rm vector}}+V_{\rm{M}}+V_{\rm{S B}}, \notag \\
&V_{{\rm vector}}=-\frac{1}{2} m_{\omega}^{2} \omega^{2}-\frac{1}{2} m_{\rho}^{2} \rho^{2}, \notag \\
&V_{\rm{M}}=-\frac{\bar{\mu}^{2}}{2} \sigma^{2}+\frac{\lambda_{4}}{4} \sigma^{4}-\frac{\lambda_{6}}{6} \sigma^{6}-\frac{1}{2}\sigma^{2} \mu_{I}^{2}\sin^2{\pi}, \notag \\
&V_{\rm{S B}}=-f_{\pi} \sigma m_{\pi}^{2} \cos \pi. 
\end{align}

\subsection{Determination of model parameters}

The present model includes following
12 relevant parameters:
\begin{align}
& m_0\,,\ g_1\,,\ g_2\,,\ g_{\omega NN} \,,\ g_{\rho NN} \,,\ \bar{\mu}^2 \,,\ \lambda_4 \,,\ \lambda_6 \,, \notag\\
& m_\omega \,,\ m_\rho \,,\ m_\pi \,,\ f_\pi \ .
\end{align}
First of all, we use the physical values for $m_\omega$, $m_\rho$, $m_\pi$ and  $f_\pi$ as shown in Table~\ref{tab:masses}.
\begin{table}[htbp]
    \centering
    \begin{tabular}{cccc}
       \hline  $m_\omega[\mathrm{MeV}]$  & $m_\rho[\mathrm{MeV}]$ & $m_\pi[\mathrm{MeV}]$ & $f_\pi[\mathrm{MeV}]$ \\ \hline 783
         &776 &140 & 92.4 \\ \hline
    \end{tabular}
    \caption{
    Masses and the pion decay constant used in the analysis.
    }
    \label{tab:masses}
\end{table}
The scalar coupling constants 
$g_1$ and $g_2$ are determined by using $m_+=939$ MeV and $m_-=1535$ MeV as inputs for given value of the chiral invariant mass $m_0$.
The results are shown in Table \ref{tab:yukawacoupling}.
\begin{table}[h]
\centering
$\begin{array}{c|ccccc}
\hline m_{0}[\mathrm{MeV}] & 500 & 600 & 700 & 800 & 900 \\
\hline g_{1} & 9.02 & 8.48 & 7.81 & 6.99 & 5.96 \\
g_{2} & 15.47 & 14.93 & 14.26 & 13.44 & 12.41 \\
\hline
\end{array}$
\caption{Values of scalar Yukawa coupling constants for different chiral invariant mass}
\label{tab:yukawacoupling}
\end{table}

Because pion condensate does not occur at nuclear saturation density in symmetric nuclear matter, so that 
we simply use the method introduced in 
Ref.~\cite{Motohiro:2015taa}, and 
determine the remaining five  
parameters by fitting
to the nuclear saturation properties
shown in Table~\ref{tab:saturationinputs},
combined with the stationary condition for the meson potential at vacuum which is expressed as
\begin{equation}
-\bar{\mu} f_\pi + \lambda_4 f_\pi^3 - \lambda_6 f_\pi^5 - m_\pi^2 f_\pi = 0 \, .
\end{equation}

\begin{table}[H]
    \centering
    \begin{tabular}{cccc}
    \hline
    $n_0\,[\mathrm{fm^{-3}}]$ & $B_0\,[\mathrm{MeV}]$ & $K_0\,[\mathrm{MeV}]$ & $S_0\,[\mathrm{MeV}]$ \\
    \hline
    0.16 & 16 & 240 & 31 \\
    \hline
    \end{tabular}
    \caption{Saturation properties of normal nuclear matter. $n_0$, $B_0$, $K_0$ and $S_0$ are the saturation density, the biding energy, the incompressibility and the symemtry energy.}
    \label{tab:saturationinputs}
\end{table}
The resultant parameters are shown in Table~\ref{tab:modelpar}.
\begin{table}[
h
]
\centering
    $\begin{array}{c|ccccc}
\hline m_{0}[\mathrm{MeV}] & 500 & 600 & 700 & 800 & 900 \\
\hline \sigma_{0}[\mathrm{MeV}] & 46.981 & 56.126 & 60.552 & 61.124 & 57.991 \\
\omega_0[\mathrm{MeV}] & 22.748 & 18.298 & 14.641 & 11.346 & 7.060\\
\bar{\mu}^{2} / f_{\pi}^{2} & 22.701 & 22.350 & 19.278 & 11.930 & 1.498 \\
\lambda_{4} & 41.940 & 40.392 & 35.460 & 23.118 & 4.429 \\
\lambda_{6} f_{\pi}^{2} & 16.943 & 15.746 & 13.887 & 8.893 & 0.636 \\
g_{\omega N N} & 11.345 & 9.125 & 7.302 & 5.658 & 3.521 \\
g_{\rho N N} & 7.316 & 7.855 & 8.131 & 8.298 & 8.436 \\
\hline
\end{array}$
\caption{Values of model parameters and the mean fields at saturation, $\sigma_0$ and $\omega_0$.
}
\label{tab:modelpar}
\end{table}
%%%%%%%%%%%%%%%%%%%%%%%%%%%%%%

\section{Construction of nuclear matter with pion condensate}
\label{sec:matter}
\subsection{Thermodynamic potential}
To get the thermodynamic potential of the system, path integral approach is a powerful tool to use. Here we use the standard process to calculate the thermodynamic potential. The partition function in the imaginary-time formalism is given by
\begin{equation}
    Z=\int \prod_{i=1,2}^{}  [dN_i][dN^\dagger_i]\ \exp\ (\int_{0}^{\beta}\mathrm{d}\tau\int \mathrm{d}^3x \ \mathcal{L}).
\end{equation}
with the Lagrangian in matrix form
\begin{equation}
\mathcal{L}={N}_i^\dagger D_nN_i,
\end{equation}
The matrix $D_n$ in momentum space is given by
\begin{equation}
    D_n(P)=(i\omega_n-\mathcal{H}(p)),
\end{equation}
Then the partition function is calculated as
\begin{equation}
    \ln Z = \int \frac{ d^3 p }{(2\pi)^3 } T \sum_n \mbox{tr} \ln D_n (p).
    \label{eq:partitionfunc}
\end{equation}
In the above expression, $\mathcal{H}(p)$ is the Hamiltonian in the momentum space and $\omega_n$ is the Matsubara frequency $\omega_n=(2n+1) \pi T$  ($n$ is an integer). We note  that $D_n(p)$ is the matrix in the space of the  parity doubling, the isospin and the spinor spaces.
The trace in Eq.\eqref{eq:partitionfunc} is taken in these spaces. By diagonalizing the Hamiltonian, the partition function $Z$ is calculated as
\begin{equation}
    \ln Z = \int \frac{d^3 p}{ (2\pi)^3 } \sum_i I (E_i)
\end{equation}
where
\begin{equation}
I (E_i) = T \sum_n \ln ( i \omega_n - E_i ),
\end{equation}
with $E_i$ being the $i$th eigenvalue of the Hamiltonian ${\mathcal H}(p)$.
Performing the Matsubara frequency sum, we finally get the thermodynamic potential as
\begin{equation}
\Omega_{\mathrm{PDM}}=\Omega_{\mathrm{nucleon}}+\Omega_{\mathrm{meson}}\, ,
\end{equation}
\begin{align}
\Omega_{\rm nucleon} & =  -P = -\frac{1}{\beta V}\ln Z \notag\\
& =-\int \frac{\mathrm{d}^{3} k}{(2 \pi)^{3}} \sum_{i} \frac{1}{2}\left|E_{i}\left(\mu_{B}, \mu_{I}, \phi \right)\right|\, .
\end{align}

\subsection{Subtraction scheme}

In the last 
subsection, we formaly obtained the thermodynamic potential for nucleons. To make the thermodynamic potential finite, we adopt the following subtraction scheme:
\begin{align}
& \Omega_{{\rm nucleon}} = \notag\\
& \quad -\int \frac{\mathrm{d}^{3} k}{(2 \pi)^{3}} \sum_{i}\left(\frac{1}{2}\left|E_{i}\left(\mu_{B}, \mu_{I}, \phi \right)\right|-\frac{1}{2}\left|E_{i}(0,\mu_I, \phi)\right|\right),
\label{subtracted Omega}
\end{align}
where $\phi = (\sigma, \omega, \rho, \pi)$ represents the mean field values of mesons.
In the following, we explain why we subtract $E_i(0,\mu_I,\phi)$ instead of $E_i(0,0,\phi)$. Actually, if we subtracted $E_i(0,0,\phi)$, a divergence would arise from the momentum integration. This implies that subtracting $E_i(0,0,\phi)$ is not enough in our case. 
The reason why we include the chemical potential in the subtraction term as $E_i(0,\mu_I,\phi)$ is that proton and neutron carry different chemical potentials:
\begin{align}
\mu_{p}^{*} & =\mu_{B}^*+\frac{1}{2}\mu_{I}^{*}, \notag \\
\mu_{n}^{*} & =\mu_{B}^*-\frac{1}{2}\mu_{I}^{*}.
\end{align}
where the effective baryon and isospin chemical potentials are defined as
\begin{align}
\mu_{B}^{*} & =\mu_{B}-g_{\omega N N} \omega, \notag \\
\mu_{I}^{*} & =\mu_{I}-g_{\rho N N} \rho.
\end{align}
To understand and seek a better subtraction 
scheme, we first look at the Lagrangian (\ref{eq:finallagrangian}).
It is obvious that all the pion mean fields are coupled 
to the nucleon with the isospin chemical potential $\mu_I$. If we set $\mu_I=0$, pion fields are unaffected by the nucleons.
On the other hand, when calculating the first term $|E_i(\mu_B,\mu_I,\phi)|$, the coupling of the pion field to the nucleons introduces effects that include divergences. Consequently, it is evident that simply subtracting $|E_i(0,0,\phi)|$ is insufficient to eliminate these divergences. In Appendix~\ref{appendix:sub}, we show that keeping $\mu_I^{\ast}$ is enough to subtract the divergence by taking the one-loop correction to the two-point function of $\pi$ mean field as an example.

\subsection{Gap equations}
Now we are going to compute the pressure as a function of baryon number chemical potential $\mu_B$, isospin chemical potential $\mu_I$.
For given values of $\mu_B$ and $\mu_I$, four 
mean field values $\sigma$, $\pi$, $\omega$ and $\rho$ are determined by solving gap equations from the stationary condition of the thermodynamic potential.
\begin{equation}
\label{eq:gapeqs}
    \frac{\partial \Omega_{\mathrm{PDM}}}{\partial \sigma}=\frac{\partial \Omega_{\mathrm{PDM}}}{\partial \pi}=\frac{\partial \Omega_{\mathrm{PDM}}}{\partial \omega}=\frac{\partial \Omega_{\mathrm{PDM}}}{\partial \rho}=0.
\end{equation}

In the present analysis, we numerically diagonalize the Hamiltonian given in Eq.~(\ref{hamiltonian}).
Then, instead of performing the numerical differentiation, we utilize Feynman-Hellman Theorem to solve the gap equation numerically.
\iffalse
According to the Feynman-Hellman Theorem
\begin{equation}
\frac{\mathrm{d} E_{\lambda}}{\mathrm{d} \lambda}=\left\langle\psi_{\lambda}\left|\frac{\mathrm{d} \hat{H}_{\lambda}}{\mathrm{d} \lambda}\right| \psi_{\lambda}\right\rangle
\end{equation}
here the $E_\lambda$ are the energy eigenvalues of the Hamiltonian $\hat{H}_\lambda$ (both depend on parameter $\lambda$) $\hat{H}_{\lambda}\left|\psi_{\lambda}\right\rangle=E_{\lambda}\left|\psi_{\lambda}\right\rangle$. 
\fi
%\cmh{[! I do not think that we need to refer to the general case of the F-H theorem. !]}
From the nucleon part of the thermodynamic potential given in Eq.~(\ref{subtracted Omega}), the gap equation of nucleon part is written as
\begin{align}
&\frac{\partial \Omega_{\mathrm{nucleon}}}{\partial \phi}
%\notag \\
%&
=-\int \frac{\mathrm{d}^{3} k}{(2 \pi)^{3}} \sum_{i}\Bigg[\frac{1}{2}\frac{\partial \left|E_{i}\left(\mu_{B}, \mu_{I}, \phi \right)\right|}{\partial \phi}\notag\\
& \qquad\qquad\qquad\qquad {} -\frac{1}{2}\frac{\partial\left|E_{i}(0,\mu_I, \phi)\right|}{\partial \phi}\Bigg] \ .%\notag
\end{align}
The derivatives of energy eigenvalues are expressed as
\begin{align}
\frac{\partial \left|E_i\right|}{\partial \phi} 
&= \frac{\partial}{\partial \phi} \left(E_i \Theta(E_i) - E_i \Theta(-E_i) \right) \notag \\
&= \frac{\partial E_i}{\partial \phi}\left( \Theta(E_i) - \Theta(-E_i) \right) \ .
\end{align}
Now, for evaluating this derivative numerically, we use the Feynman-Helmann theorem: 
\begin{align}
\frac{\partial E_i}{\partial \phi} = \sum_{k,j} \left(U\right)_{ij} \left( \frac{\partial \hat{H}}{\partial \phi} \right)_{jk} \left(U^{\dagger}\right)_{ki}
\ ,
\end{align}
where $U$ and $U^\dagger$ are 
the matrices that diagonalize the Hamiltonian.

\section{Phase structure}
\label{sec:phase}
\subsection{Existence of pion condensate and chiral symmetry restoration}
\begin{figure}[h]
    \centering
    \includegraphics[width=1\linewidth]{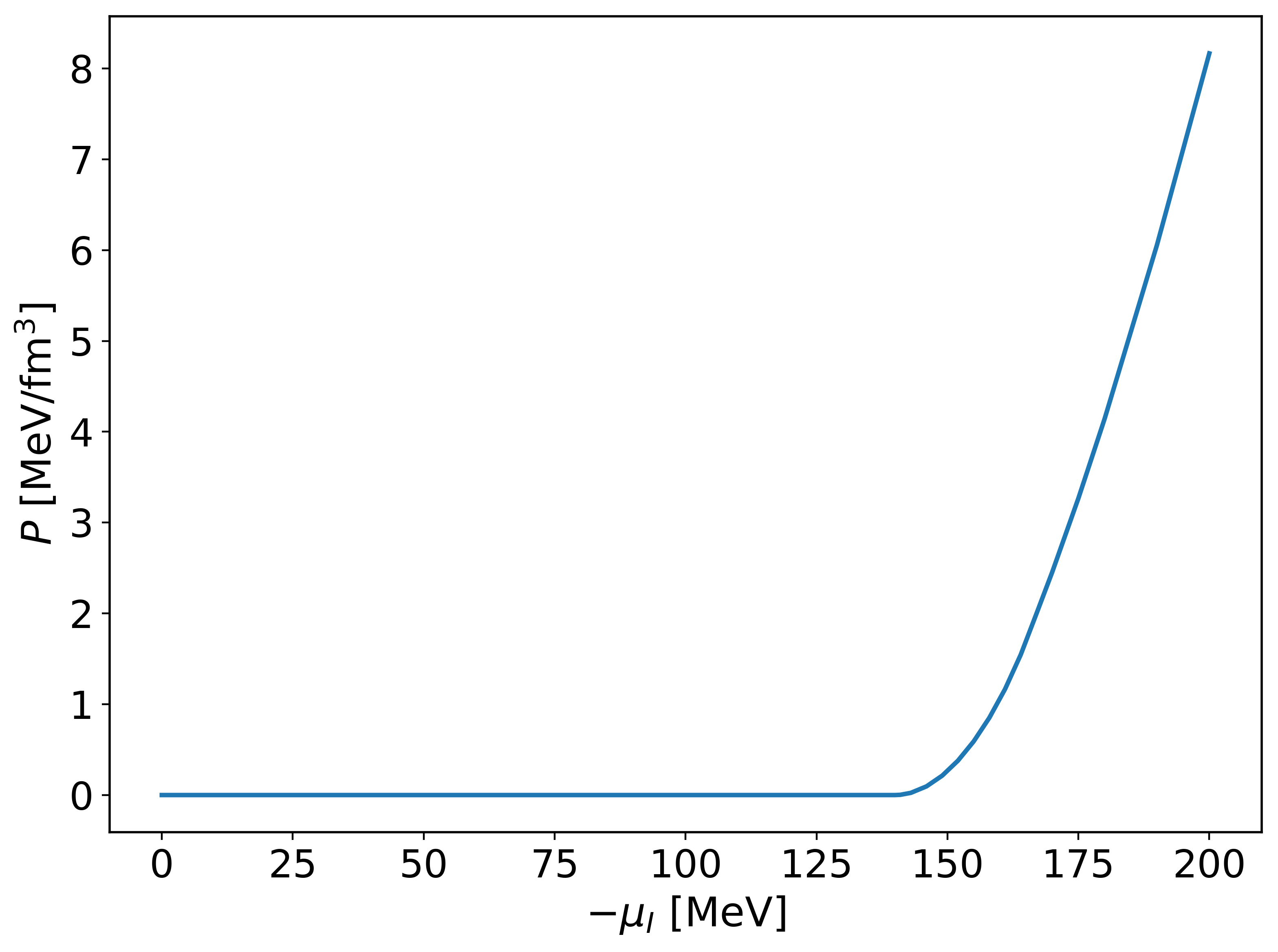}
    \caption{Contribution of pion condensate to pressure for $\mu_B=0$ and $m_0=800$\,MeV.}
    \label{fig:vacuumpressure}
\end{figure}

Let us first check the existence of the pion condensation phase in pure mesonic matter. 
In Fig.~\ref{fig:vacuumpressure},
we show the dependence of pressure $P$ on the isospin chemical potential $\mu_I$ for $\mu_B = 0$. In this figure, the pressure begins to increase from zero after the isospin chemical potential $\mu_I$ becomes larger than the critical value $m_\pi=140$ MeV. Since there is no contribution from baryons at $\mu_B=0$, this contribution comes from pion condensate. This result is consistent with the result from chiral perturbation theory \cite{Son:2000xc}. %\cmh{[! need references !]}
%\\

\begin{figure}[h]
    \centering    \includegraphics[width=1\linewidth]{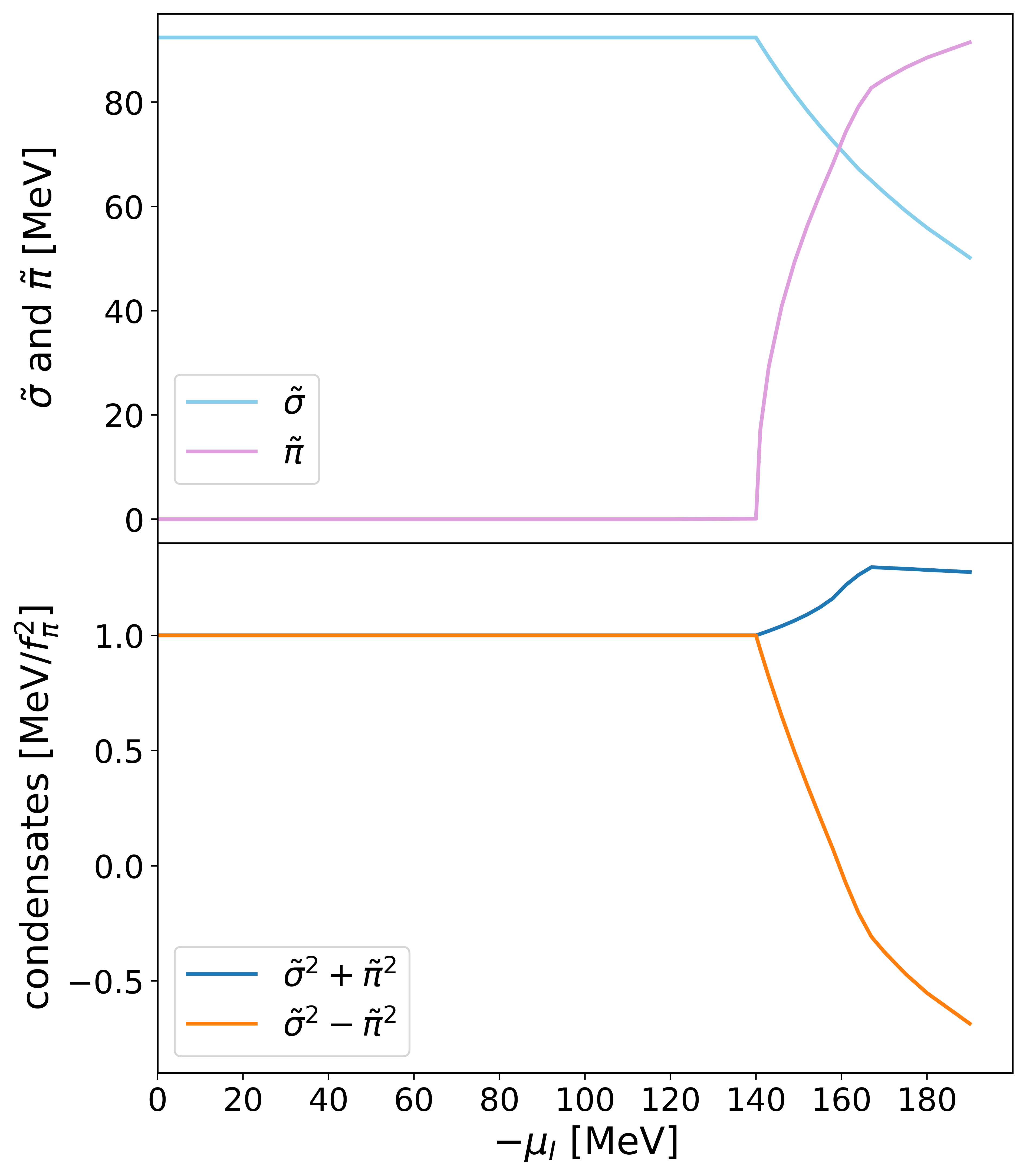}
    \caption{Mean field values for $\mu_B$=0\,MeV and $m_0=800$\,MeV. The upper panel shows $\tilde{\sigma}$ and $\tilde{\pi}$. The lower panel shows $\tilde{\sigma}^2+\tilde{\pi}^2$ and $\tilde{\sigma}^2-\tilde{\pi}^2$}
    \label{fig:sigpivac}
\end{figure}
In the upper panel of Fig.~\ref{fig:sigpivac} we show the $\tilde{\sigma}$ and $\tilde{\pi}$ mean field values at vacuum for $\mu_B$=0 and $m_0=800$\,MeV, where $\tilde{\sigma}$ and $\tilde{\pi}$ are defined as
\begin{align}
\tilde{\sigma}=\sigma\cos{\pi}, \notag \\
\tilde{\pi}=\sigma\sin{\pi}.
\end{align}
For $-\mu_I<140$\,MeV, $\tilde{\pi}=0$ implies that pion condensation does not occur. For $-\mu_I>140$\,MeV, pion condensation occurs and $\tilde{\sigma}$ is suppressed.
In the lower panel of Fig.~\ref{fig:sigpivac}, we show $\tilde{\sigma}^2+\tilde{\pi}^2$ and $\tilde{\sigma}^2-\tilde{\pi}^2$ at $\mu_B=0$. The combination $\tilde{\sigma}^2+\tilde{\pi}^2$ is invariant under chiral rotations and characterizes the magnitude of the chiral condensate, whereas $\tilde{\sigma}^2-\tilde{\pi}^2$ depends on both its magnitude and orientation in the $(\tilde{\sigma},\tilde{\pi})$ plane. As $\mu_I$ increases beyond the onset of pion condensation, $\tilde{\sigma}^2+\tilde{\pi}^2$ increases and then remains stable. On the other hand, $\tilde{\sigma}^2-\tilde{\pi}^2$ first decreases to zero and then becomes negatively large. This behavior indicates a redistribution of the condensate from the sigma direction toward the pion direction, accompanied by an increase in its total magnitude. Chiral symmetry is once restored and then broken again, with the magnitude enhanced by the pion condensate.~\footnote{This enhancement is pointed out in Ref.~\cite{Nishihara:2014nva}.]}
\begin{figure}[h]
    \centering
    \includegraphics[width=1\linewidth]{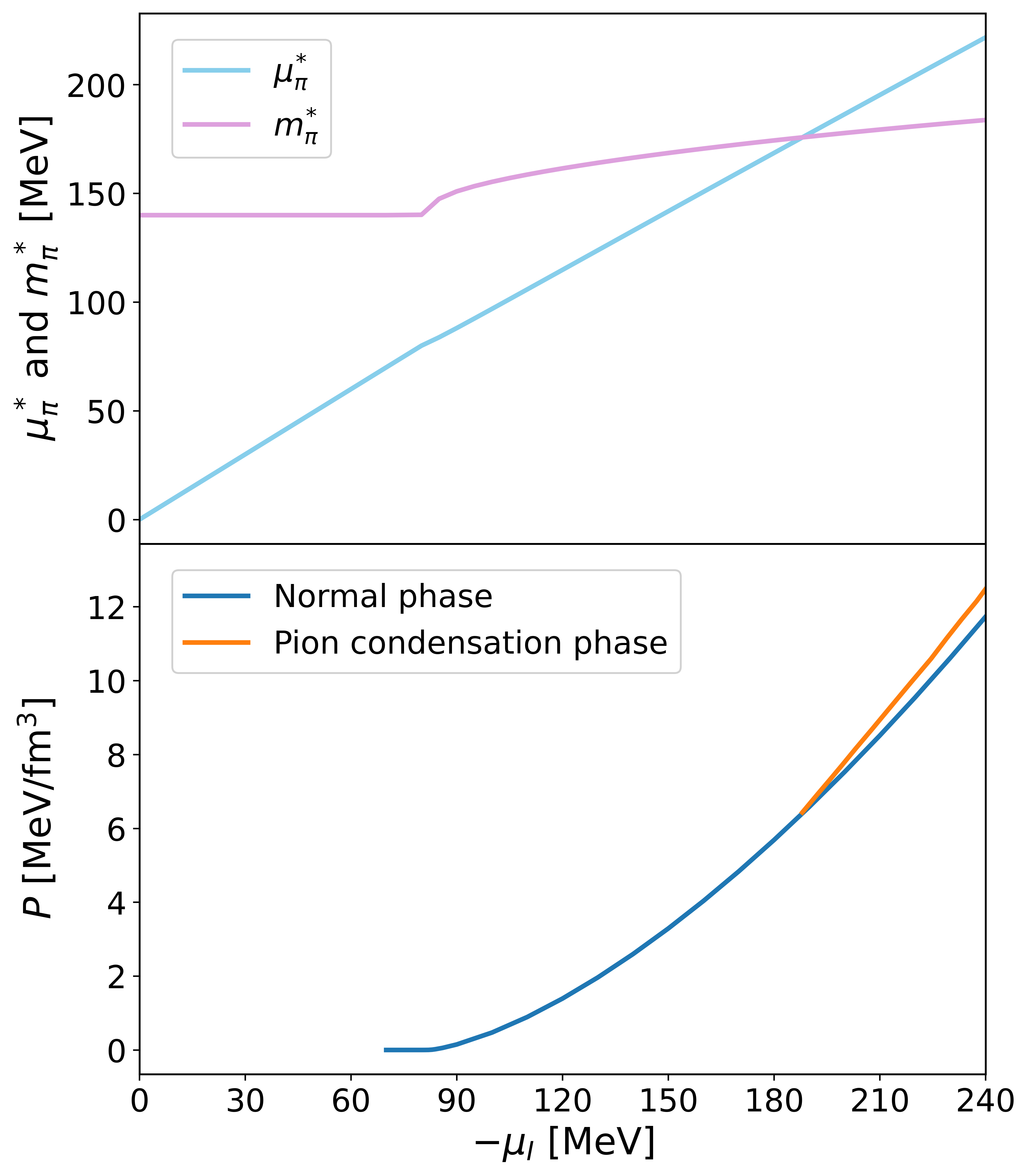}
    \caption{
    Isospin chemical potential dependences of the effective pion chemical potential $\mu_\pi^*$ and mass $m_\pi^*$ (upper panel), and the pressure (lower panel), for $\mu_B= 900$\,MeV and $m_0=800$\,MeV.
    }
    \label{fig:pressure}
\end{figure}

Next, we include the effect of nucleons. In the upper panel of Fig.~\ref{fig:pressure},
we show the effective pion chemical potential $\mu_\pi^*$ and mass $m_\pi^*$, which are defined as (see Appendix~\ref{appendix:pion} for derivation)
\begin{equation}
m_\pi^* = \sqrt{\frac{f_\pi m_\pi^2}{\sigma}}, \qquad \mu_\pi^*=\mu_I-\frac{m_\rho^2}{2g_\rho^2 \sigma^2}\rho.
\end{equation}
In the lower panel, pressure for both normal phase (blue curve) and pion condensation phase (orange curve) are given.
When $-\mu_I$ increases, both $\mu_\pi^*$ and $m_\pi^*$ increase. When $-\mu_I$ is smaller than the critical point ($-\mu_I < -\mu_I^{\text{cr}}= 188\,$MeV), $m_\pi^*$ is larger than $\mu_\pi^*$ ($m_\pi^*>\mu_\pi^*$) and pion condensation does not occur. At the critical point $-\mu_I = -\mu_I^{\text{cr}}$, $m_\pi^*$ agrees with $\mu_\pi^*$. For $-\mu_I > -\mu_I^{\text{cr}}$, $\mu_\pi^*>m_\pi^*$ and we obtain two solutions by solving the gap equations~(\ref{eq:gapeqs}): one gives $\pi=0$ and the other gives a non-zero $\pi$. As shown in Fig.~\ref{fig:pressure}, the pressure of the normal phase (blue curve) with $\pi=0$ is smaller than that of the pion condensation phase (orange curve) with $\pi\neq0$. This implies that the pion condensation phase is realized for $-\mu_I > -\mu_I^{\text{cr}}$. From the upper panel of Fig.~\ref{fig:sigpivac}, this phase transition is of second order.

\subsection{Phase diagram and dependence on $m_0$}
\begin{figure}[h]
    \centering
    \includegraphics[width=1\linewidth]{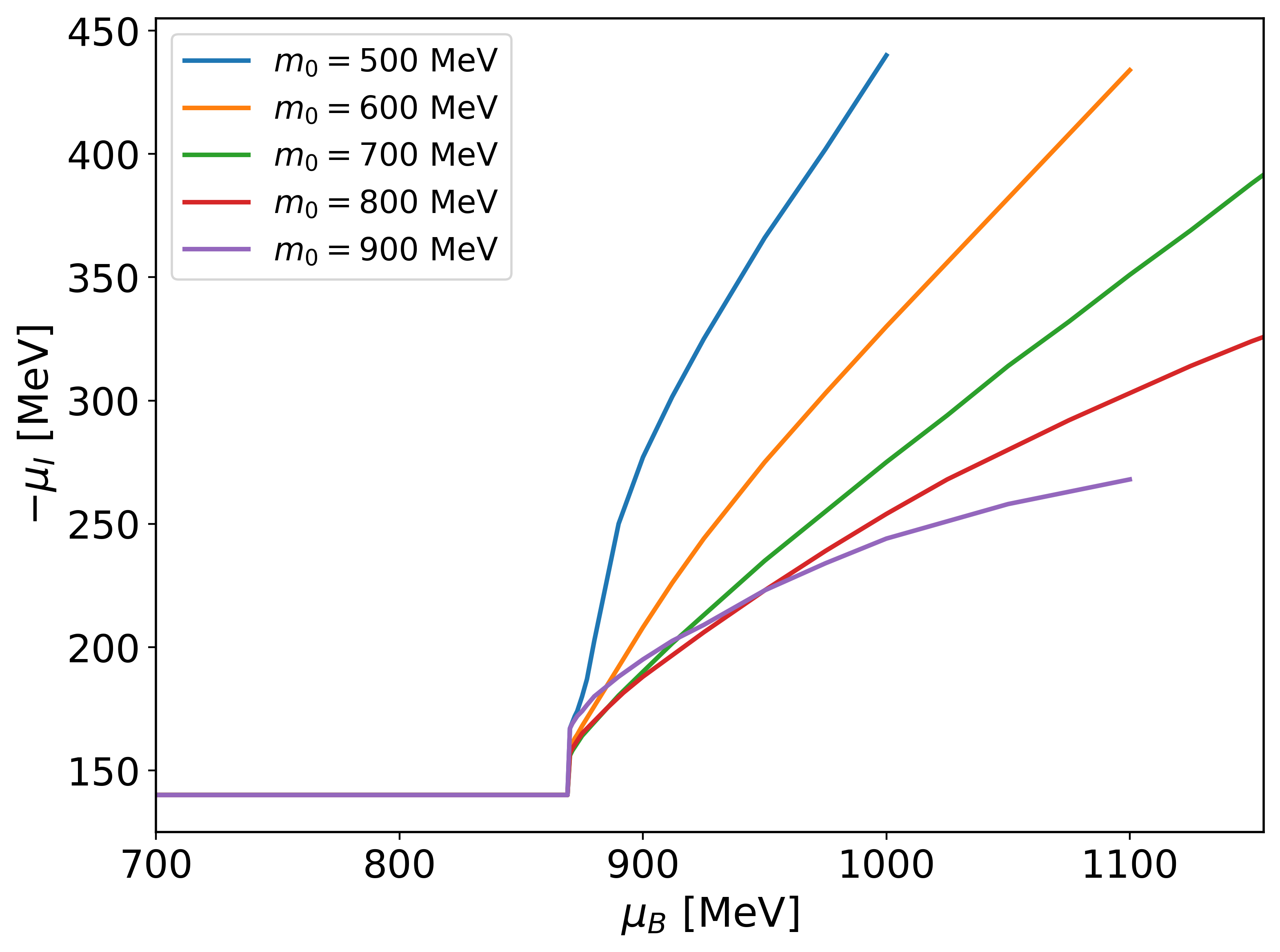}
    \caption{
    Phase boundary for different $m_0$}
    \label{fig:phasetransline}
\end{figure}

In Fig.~\ref{fig:phasetransline}, we show the $m_0$ dependence of phase boundary between normal phase and the pion condensation phase on the $\mu_I-\mu_B$ plane. For $\mu_B < 869$ MeV, the phase boundary—traced by increasing $\mu_I$ at a fixed $\mu_B$—coincides at the crtical $\mu_I$, $-\mu_I^{\text{cr}} = 140$ MeV for all values of $m_0$.
In contrast, for $\mu_B > 869$ MeV, $\mu_I^{\rm cr}$ found by increasing $|\mu_I|$ at a fixed $\mu_B$ exceeds $140$\,MeV and varies depending on $m_0$.
This shift of $\mu_I^{\rm cr}$ arises from baryonic effects.
It should be emphasized here that, regardless of the value of $m_0$, this change of $\mu_I^{\rm cr}$ from 140 MeV appears for $\mu_B > 869$ MeV.
The reason for this can be explained as follows:
As $\mu_B$ increases from 0 on the phase boundary, the neutron effective chemical potential $\mu_n^*$ increases as well. Since phase transition to the pion condensation phase is of second order, the pion condensation vanishes at phase boundary. Furthermore, $\omega$ and $\rho$ mean fields also vanish because there is no nucleon in matter. As a result, $\mu_n^*=\mu_n=\mu_B-\frac{1}{2}\mu_I$ at the phase boundary.
In addition, neutron mass is equal to the vacuum one, $m_+=939$ MeV. As $\mu_n$ increases further and reaches $m_+$, neutrons start to appear. This point is characterized by:
\begin{equation}
\mu_n^\text{cr}=\mu_{B}^\text{cr}-\frac{1}{2} \mu_{I}^\text{cr}=m_{+},
\end{equation}
where  $-\mu_I^\text{cr}=m_\pi=140$ MeV. This determines $\mu_B^\text{cr}=869$ MeV.

Now let us explain how baryon affects the phase boundary. We can see from Fig.~\ref{fig:phasetransline}, after baryon enters into the matter, phase boundary starts to bend upward, which means that $-\mu_I > m_\pi$ is needed to induce the pion condensation. The reason why the existence of baryon makes the critical $\mu_I$ larger than $m_\pi$ is as follows:
%\cmh{[! The word ``delay'' is hard to understand.  Please put explanation in more detail. !]}
Before baryon enters, all the energy from  isospin chemical potential is used to induce pion. 
After baryon enters into the matter, part of the energy is used to induce baryons and so that we need more energy (a larger isospin chemical potential) to induce pion in this case. As a result, the phase boundary bends upward and the $\mu_I^\text{cr}$ is increased.

\begin{figure*}[htp]
    \centering
\includegraphics[width=1\linewidth]{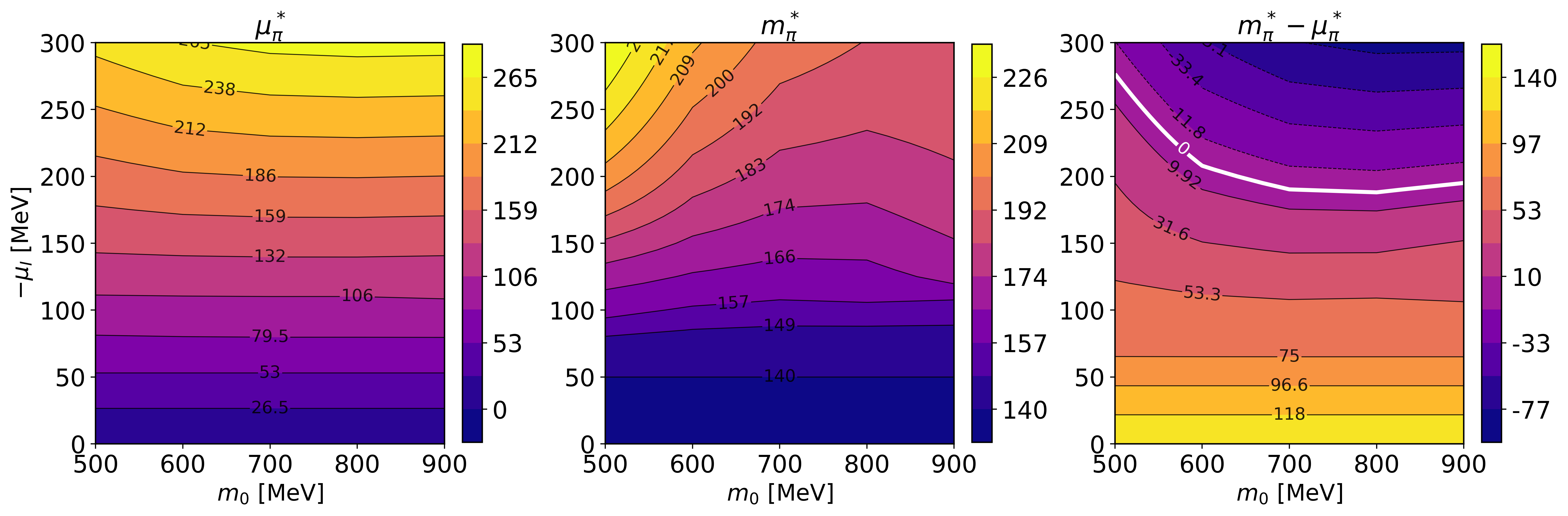}
    \caption{
    Dependence of
    $\mu_\pi^*$, $m_\pi^*$ and $m_\pi^*-\mu_\pi^*$
    on $m_0$ and $-\mu_I$.
    }
    \label{fig:effective_values}
\end{figure*}

We now turn our attention on $m_0$ dependence of the phase boundary.
In Fig.~\ref{fig:phasetransline}, for a fixed $\mu_B$, $\mu_I^\text{cr}$ is larger for a smaller $m_0$. We explain this by studying the tree-level effective pion chemical potential $\mu_\pi^*$ and effective pion mass $m_\pi^*$.
In fig.~\ref{fig:effective_values}, we show dependences of $\mu_\pi^*$ (left panel), $m_\pi^*$ (middle panel) and $m_\pi^*-\mu_\pi^*$ (right panel) on $m_0$ and $\mu_I$, respectively.
The left panel shows that $\mu_\pi^*$ depends on $\mu_I$ but is not sensetive to $m_0$, since $\frac{m_\rho^2}{2g_\rho^2 \sigma^2}\rho$ is much smaller than $\mu_I$. While the middle panel shows that $m_\pi^*$ depends on both. For fixed $\mu_I$, $m_\pi^*$ is larger for smaller $m_0$
excluding the region where $m_0 = 800$ - $900$\,MeV.
This can be understood as follows: The Yukawa coupling $g_1$ and $g_2$ are larger for smaller $m_0$, which leads to larger correction to $\sigma$ resulting smaller $\sigma$ for smaller $m_0$. In the right panel, the white curve shows the phase boundary where $m_\pi^*=\mu_\pi^*$, which means $\mu_I^\text{cr}$ is larger for smaller $m_0$.

\subsection{Liquid-Gas phase transition}
\label{subsec:LG}

In this subsection, we study how the phase boundary of the liquid-gas phase transition is affected by the pion condensation.
\begin{figure}[h]
    \centering
    \includegraphics[width=1\linewidth]{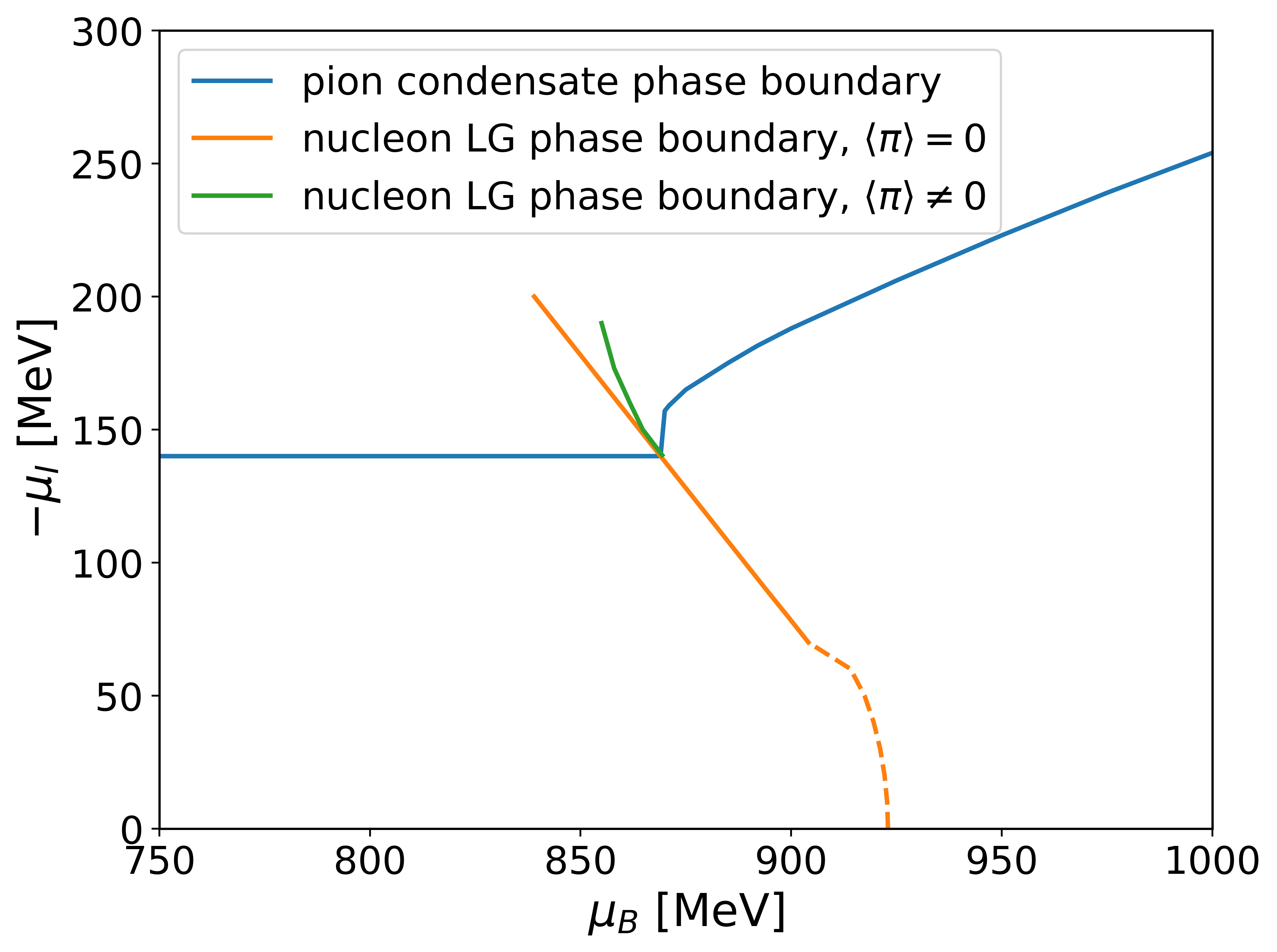}
    \caption{Liquid-Gas phase transition line with pion condensation for $m_0=800$ MeV 
    }
    \label{fig:liquid_gas}
\end{figure}
In Fig.~\ref{fig:liquid_gas},
we show the Liquid-Gas coexistence curves. The orange curve from $(\mu_B,-\mu_I)=(923, 0)$ to $(\mu_B,-\mu_I)=(839, 200)$
shows the Liquid-Gas coexistence curve drawn by assuming no pion condensation. For $-\mu_I<65$\,MeV, the Liquid-Gas phase transition is of first order (dashed curve), while for $-\mu_I>65$\,MeV, it is of second order (solid curve). 
For $-\mu_I > m_\pi=140$\,MeV, pion condensate occurs, the green curve shows the coexisting curve with the effect of pion condensation included. With the onset of pion condensation, the green curve departs from orange curve and starts to bend upward.
As a result, for a fixed $-\mu_I$, the critical baryon chemical potential $\mu_B^\text{cr}$ for the Liquid-Gas phase transition becomes larger when pion condensation is considered. The reason is as below: in the present system, the neutron first enters to the matter. Part of the energy (isospin chemical potential) is used to populate pion, which implies that a larger $\mu_B$ is needed to populate neutron.

\section{Comparison with NS matter}

In this section, we investigate whether pion condensation occurs inside neutron stars.
\label{sec:NS}
\begin{figure}[h]
    \centering
    \includegraphics[width=1\linewidth]{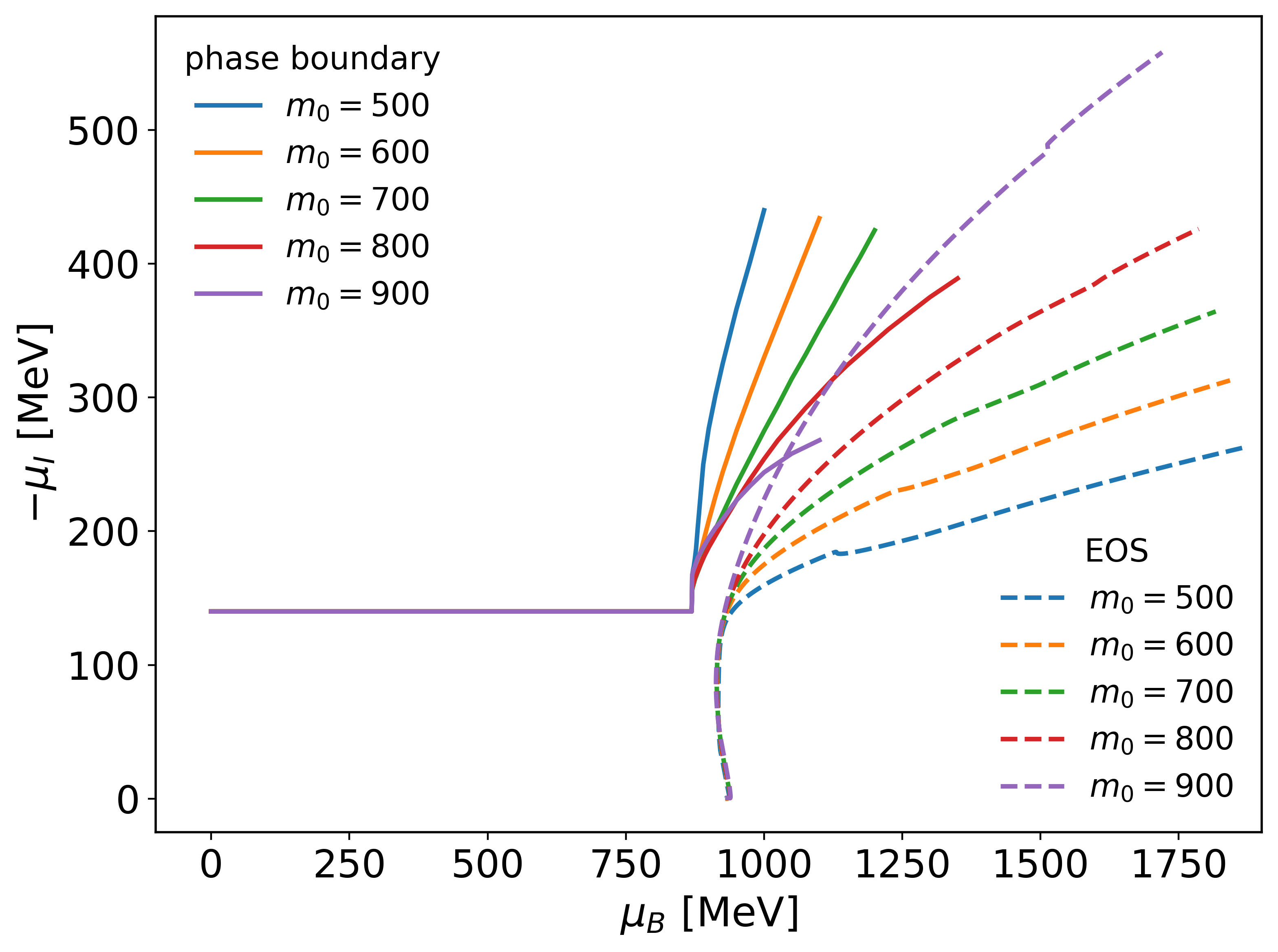}
    \caption{
    Relations between $\mu_B$ and $\mu_I$ in the neutron star matter (dashed curves)
    together with the phase boundaries for the pion condensation (solid curves).
    }
    \label{fig:NSmatter}
\end{figure}
In Fig.~\ref{fig:NSmatter}, the dashed curves show the relation between the baryon number chemical potential $\mu_B$ and the isospin chemical potential $\mu_I$ realized inside the neutron stars, which is determined by imposing $\beta$-equilibrium and electric charge neutrality with assuming no pion condensation. Solid curves are the phase boundaries as shown in Fig.~\ref{fig:phasetransline}.
We see only the purple curves (phase boundary and the relation between $\mu_B$ and $\mu_I$ inside neutron star matter for $m_0=900$MeV) in 
Fig.~\ref{fig:NSmatter} intersects with each other, which indicates that pion condensate appears in neutron star matter at this point and the equation of state should be modified for density higher than this.
However, we notice that at this intersection point, the density is already near four times of 
nuclear saturation density $n_0$, where the parity doublet model may not be applicable. On the other hand, if we use pure PDM matter to solve Tolman-Oppenheimer-Volkoff (TOV) equation to obtain the Mass-Radius relation, it is difficult to satisfy the two-time solar mass constraint. Therefore we conclude that in the present parity doublet model, pion condensate is unlikely to occur in neutron star matter.

\section{SUMMARY AND DISCUSSIONS}
\label{sec:summary}

In this work, we constructed nuclear matter based on the
parity doublet model proposed in Ref.~\cite{Motohiro:2015taa},
taking pion condensate into account.
We proposed a new subtraction scheme to obtain the finite thermodynamic potential within the no-sea approximation.
Differently from the standard
no-sea approximation, we keep $\mu_I^*$ in the subtraction part while we set $\mu_B^*$ to be zero.
After obtaining the thermodynamic potential, we solved the four gap equations for a given $\mu_I$ and $\mu_B$ to determine the four mean fields, $\sigma$, $\omega$, $\rho$ and $\pi$.
We found that pion condensate actually occurs when $\mu_I$ is large enough. However, the existence of baryons makes
the critical isospin chemical potential $\mu_I^{\rm cr}$ larger than $m_\pi = 140$\,MeV.
The mechanism of this phenomenon is explained as follows: When $\mu_n = \mu_B - \frac{1}{2} \mu_I$ is smaller than the neutron mass, all the energy
from isospin chemical potential is used to induce pion, so that the pion condensate occurs when $- \mu_I > m_\pi$. 
For $\mu_n > m_n$, on the other hand, part of the energy is used to induce neutron and the pion condensate is suppressed.

We studied the $m_0$ dependence of phase boundary between the pion condensation phase and the normal phase. Our results showed that, 
for a fixed $\mu_B$, $\mu_I^{\rm cr}$ is larger for a smaller $m_0$,
which is mainly caused by the $m_0$ dependence of the effective pion mass $m_\pi^\ast$. We also studied the effect of the pion condensation to the Liquid-Gas coexistence curve of nuclear matter. The result shows that for a given $\mu_I$, the critical baryon chemical potential $\mu_B^\text{cr}$ for the Liquid-Gas phase transition becomes larger when pion condensation occurs. This is because part of the energy (chemical potential) used to induce neutron is now employed to induce pion condensate.

We compared the resultant phase boundary for the pion condensate on the phase diagram
with the relation of $\mu_B$ and $\mu_I$ in the neutron star matter.
Our result shows that there is a possibility of the appearance of the pion condensate in the neutron star matter for $m_0 = 900$\,MeV, but the density is four times of normal nuclear density. Thus we conclude that pion condensate is unlikely to occur in neutron star in the present parity doublet model.

For future work, including $\omega$-$\rho$ mixing term is one thing valuable to do. The motivation can be explained from two perspectives. One is that adding $\omega$-$\rho$ mixing helps to reduce the slope parameter $L_0$. In current model, the slope parameter at nuclear saturation density is around $80$\,MeV. However, this values 
is a little large if we compared with the empirical value $L_0= 57.7 \pm 19$\,MeV shown in
Ref.~\cite{Li:2021thg}.
%B.-A. Li, B.-J. Cai, W.-J. Xie, and N.-B. Zhang, ``Progress in constraining nuclear symmetry energy using neutron star ob- servables since GW170817,'' Universe 7, 182 (2021).
Besides, according to Refs.~\cite{Gao:2022klm,Kong:2023nue},
$\omega$-$\rho$ mixing term can soften the equation of states of neutron star.
On the phase diagram, the possibility  of intersection increases when EoSs shift to the left.

We can also extend our model to three-flavor nuclear matter to include hyperons and kaon condensate. In particular, a phase composed of kaon condensate and hyperon mixed baryonic matter is shown to exist in neutron star (see, e.g. Refs.~\cite{Muto:2024upf, Bhat:2026hco}). The phase was never studied in the parity doublet models, to our best knowledge. It is interesting to study it bt extending the parity doublet model to three-flavor as done in Refs.~\cite{Gao:2024mew,Minamikawa:2023ypn,Gao:2026scv,Gao:2026jsg}.
Additionally, it is interesting to study the effect of meson condensates to the neutron star cooling (see, e.g. Refs.~\cite{Bhat:2026hco, Brodie:2025nww, Negreiros:2026ode}).

%%%%%%%%%%%%%%%%%%
\section*{Acknowledgement}
The work of X. Liu is financially supported by JST
SPRING, Grant Number JPMJSP2125. X. Liu would
also like to use the chance to thank the “Tokai Higher
Education and Research System Make New Standards
Program for the Next Generation Researchers”. The work of  M.H. is supported in part by JSPS KAKENHI Grant Nos.~23H05439 and 24K07045. The work of B. Gao is supported by Grants-in-Aid for Scientific Research No. 26K17147.

\appendix

%\onecolumngrid

\begin{widetext}

%\cmh{[! I have changed the layout below. !]}

\section{Hamiltonian}

By performing the Legendre transformation from the Lagrangian in Eq.\eqref{eq:finallagrangian}, the baryon contribution to the Hamiltonian is expressed as
\begin{equation}
\left(\begin{array}{cc}
N_{1}^{\dagger} & {N}_{2}^{\dagger}
\end{array}\right) {\mathcal H}(p)  \left(\begin{array}{c}
N_{1} \\ N_{2}
\end{array}\right) =
\left(\begin{array}{cc}
N_{1}^{\dagger} & {N}_{2}^{\dagger}
\end{array}\right)\left( \begin{array}{cc}
A & m_0\gamma^0\gamma^5\\
-m_0\gamma^0\gamma^5  & B
\end{array} \right)\left(\begin{array}{c}
N_{1} \\ N_{2}
\end{array}\right),
\end{equation}
where
\begin{align}
A &= \gamma^{0} \bm{p}\cdot\bm{\gamma}+g_{1}\sigma_{0}\gamma^0-\left(\mu_{B}-g_{\omega N N} \omega_{0}\right)\gamma^0\gamma^0 - \left(\mu_{I} \cos \pi-g_{\rho N N} \rho_{0}\right) T^{3}\gamma^0\gamma^0 -\left(\mu_{I} \sin \pi\right) T^{2} \gamma^{0}\gamma^{0}\gamma^{5}, \notag \\
B &= \gamma^{0} \bm{p}\cdot\bm{\gamma}+g_{2}\sigma_{0}\gamma^0-\left(\mu_{B}-g_{\omega N N} \omega_{0}\right)\gamma^0\gamma^0 - \left(\mu_{I} \cos \pi-g_{\rho N N} \rho_{0}\right) T^{3}\gamma^0\gamma^0 +\left(\mu_{I} \sin \pi\right) T^{2} \gamma^{0}\gamma^{0}\gamma^{5}.
\notag
\end{align}
%}
From the rotational symmetry, we can take the first and second component of $p$ as zero. As a result, the explicit form of the matrix between $(N_1^\dagger, N_2^\dagger)$ and $(N_1, N_2)^T$ is expressed as
\begin{equation}
%\large
\left(\begin{array}{cccccccccccccccc}
a & 0 & p & 0 & 0 & 0 & \mu & 0 & 0 & 0 & m_{0} & 0 & 0 & 0 & 0 & 0 \\
0 & a & 0 & - p & 0 & 0 & 0 & \mu & 0 & 0 & 0 & m_{0} & 0 & 0 & 0 & 0 \\
p & 0 & b & 0 & \mu & 0 & 0 & 0 & -m_{0} & 0 & 0 & 0 & 0 & 0 & 0 & 0 \\
0 & - p & 0 & b & 0 & \mu & 0 & 0 & 0 & -m_{0} & 0 & 0 & 0 & 0 & 0 & 0 \\
0 & 0 & -\mu & 0 & c & 0 & p & 0 & 0 & 0 & 0 & 0 & 0 & 0 & m_{0} & 0 \\
0 & 0 & 0 & -\mu & 0 & c & 0 & - p & 0 & 0 & 0 & 0 & 0 & 0 & 0 & m_{0} \\
-\mu & 0 & 0 & 0 & p & 0 & d & 0 & 0 & 0 & 0 & 0 & -m_{0} & 0 & 0 & 0 \\
0 & -\mu & 0 & 0 & 0 & - p & 0 & d & 0 & 0 & 0 & 0 & 0 & -m_{0} & 0 & 0 \\
0 & 0 & -m_{0} & 0 & 0 & 0 & 0 & 0 & a^{\prime} & 0 & p & 0 & 0 & 0 & -\mu & 0 \\
0 & 0 & 0 & -m_{0} & 0 & 0 & 0 & 0 & 0 & a^{\prime} & 0 & - p & 0 & 0 & 0 & -\mu \\
m_{0} & 0 & 0 & 0 & 0 & 0 & 0 & 0 & p & 0 & b^{\prime} & 0 & -\mu & 0 & 0 & 0 \\
0 & m_{0} & 0 & 0 & 0 & 0 & 0 & 0 & 0 & - p & 0 & b^{\prime} & 0 & -\mu & 0 & 0 \\
0 & 0 & 0 & 0 & 0 & 0 & -m_{0} & 0 & 0 & 0 & \mu & 0 & c^{\prime} & 0 & p & 0 \\
0 & 0 & 0 & 0 & 0 & 0 & 0 & -m_{0} & 0 & 0 & 0 & \mu & 0 & c^{\prime} & 0 & - p \\
0 & 0 & 0 & 0 & m_{0} & 0 & 0 & 0 & \mu & 0 & 0 & 0 & p & 0 & d^{\prime} & 0 \\
0 & 0 & 0 & 0 & 0 & m_{0} & 0 & 0 & 0 & \mu & 0 & 0 & 0 & - p & 0 & d^{\prime} 
\end{array}\right),
\label{hamiltonian}
\end{equation}
where $a, b, c, d, a^{\prime}, b^{\prime}, c^{\prime}, d^{\prime}, $and$ \ \mu$ denote
\begin{align}
a &= -\left(\mu_{B}-g_{\omega N N} \omega_{0}\right) - \frac{1}{2}\left(\mu_{I} \cos \pi-g_{\rho N N} \rho_{0}\right) + g_1\sigma_0, \notag \\
b &= -\left(\mu_{B}-g_{\omega N N} \omega_{0}\right) - \frac{1}{2}\left(\mu_{I} \cos \pi-g_{\rho N N} \rho_{0}\right) - g_1\sigma_0, \notag \\
c &= -\left(\mu_{B}-g_{\omega N N} \omega_{0}\right) + \frac{1}{2}\left(\mu_{I} \cos \pi-g_{\rho N N} \rho_{0}\right) + g_1\sigma_0, \notag \\
d &= -\left(\mu_{B}-g_{\omega N N} \omega_{0}\right) + \frac{1}{2}\left(\mu_{I} \cos \pi-g_{\rho N N} \rho_{0}\right) - g_1\sigma_0, \notag \\
a^{\prime} &= -\left(\mu_{B}-g_{\omega N N} \omega_{0}\right) - \frac{1}{2}\left(\mu_{I} \cos \pi-g_{\rho N N} \rho_{0}\right) + g_2\sigma_0, \notag \\
b^{\prime} &= -\left(\mu_{B}-g_{\omega N N} \omega_{0}\right) - \frac{1}{2}\left(\mu_{I} \cos \pi-g_{\rho N N} \rho_{0}\right) - g_2\sigma_0, \notag \\
c^{\prime} &= -\left(\mu_{B}-g_{\omega N N} \omega_{0}\right) + \frac{1}{2}\left(\mu_{I} \cos \pi-g_{\rho N N} \rho_{0}\right) + g_2\sigma_0, \notag \\
d^{\prime} &= -\left(\mu_{B}-g_{\omega N N} \omega_{0}\right) + \frac{1}{2}\left(\mu_{I} \cos \pi-g_{\rho N N} \rho_{0}\right) - g_2\sigma_0, \notag \\
\mu &= \frac{1}{2}i\mu_I\sin\pi. \notag
\end{align}
\end{widetext}

In the calculation, we devide this Hamiltonian into two 8-by-8 matrices because the energy eigenvalues of the spin up and down parts are the same. The matrix which we use is expressed as:
\begin{equation}
\mathcal{H} = \left(\begin{array}{cccccccc}
a & p & 0 & \mu & 0 & m_{0} & 0 & 0 \\
p & b & \mu & 0 & -m_{0} & 0 & 0 & 0 \\
0 & -\mu & c & p & 0 & 0 & 0 & m_{0} \\
-\mu & 0 & p & d & 0 & 0 & -m_{0} & 0 \\
0 & -m_{0} & 0 & 0 & a^{\prime} & p & 0 & -\mu \\
m_{0} & 0 & 0 & 0 & p & b^{\prime} & -\mu & 0 \\
0 & 0 & 0 & -m_{0} & 0 & \mu & c^{\prime} & p \\
0 & 0 & m_{0} & 0 & \mu & 0 & p & d^{\prime} \notag \end{array}\right),
\end{equation}
This is to be compared with the Hamiltonian in the vacuum given by
\begin{equation}
\mathcal{H}_0 = \left(\begin{array}{cccccccc}
e & p & 0 & 0 & 0 & m_{0} & 0 & 0 \\
p & f & 0 & 0 & -m_{0} & 0 & 0 & 0 \\
0 & 0 & g & p & 0 & 0 & 0 & m_{0} \\
0 & 0 & p & h & 0 & 0 & -m_{0} & 0 \\
0 & -m_{0} & 0 & 0 & e^{\prime} & p & 0 & 0 \\
m_{0} & 0 & 0 & 0 & p & f^{\prime} & 0 & 0 \\
0 & 0 & 0 & -m_{0} & 0 & 0 & g^{\prime} & p \\
0 & 0 & m_{0} & 0 & 0 & 0 & p & h^{\prime} \notag \end{array}\right),
\end{equation}
where $e, f, g, h, e^{\prime}, f^{\prime}, g^{\prime}$ and $h^{\prime}$ denote
\begin{align}
e &= g_1\sigma_0, \notag \\
f &= - g_1\sigma_0, \notag \\
g &= g_1\sigma_0, \notag \\
h &= - g_1\sigma_0, \notag \\
e^{\prime} &= g_2\sigma_0, \notag \\
f^{\prime} &= - g_2\sigma_0, \notag \\
g^{\prime} &= g_2\sigma_0, \notag \\
h^{\prime} &= - g_2\sigma_0. \notag \\
\end{align}

\section{Subtraction scheme}
\label{appendix:sub}
\begin{figure}[H]
    \centering
    \includegraphics[width=1\linewidth]{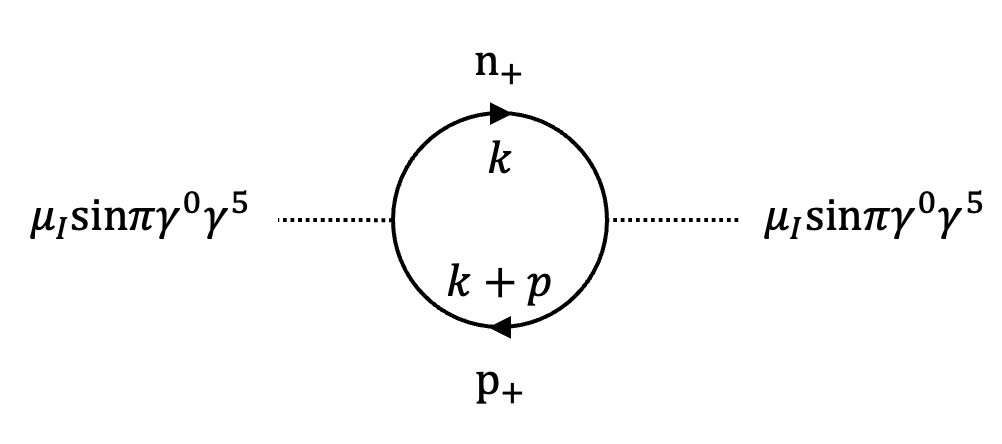}
    \caption{Feynman diagram for the one-loop contribution to the thermodynamic potential. $n_+$ ($p_+$) is the positive parity neutron (proton).
    }
    \label{fig:loop}
\end{figure}

In this section, we explain why we adopt the current subtraction scheme by considering a one-loop contribution to the thermodynamic potential as shown in Fig.~\ref{fig:loop}.
In Fig.~\ref{fig:loop}, a proton and a neutron are coupled with each other by the vertex $\mu_I\sin\pi\gamma^0\gamma^5$. We asign the zero component of momentum for the neutron as $k_0=i\omega^F_j+\mu_n^*$, and the one for the pion as $p_0=i\omega_k^B+\mu_\pi^*$,
where $\omega_j^F$ and $\omega_k^B$ are the Matsubara frequency for Fermions and Bosons, respectively. 
$\mu_n^*$ and $\mu_\pi^*$ are effective neutron and pion chemical potentials. Accordingly, the zeroth component of the proton momentum is expressed as
$k_0 + p_0 = i (\omega_j^F + \omega_k^B) + \mu_p^\ast$ with satisfying $\mu_\pi^*=\mu_p^*-\mu_n^*=\mu_I^*$. 
\begin{align}
\label{eq:loop}
\Sigma=&-T\sum_{n}\int \frac{d^3k}{(2\pi)^3} \notag \\ &\times \mathrm{Tr}[\mu_I\sin\pi\gamma^0\gamma^5\frac{m+\slashed{k}}{m^2-k^2}\mu_I\sin\pi\gamma^0\gamma^5\frac{m+\slashed{k}+\slashed{p}}{m^2-(k+p)^2}] \notag \\
&=-T\sum_{n}\int \frac{d^3k}{(2\pi)^3}4\mu_I^2 \sin^2\pi\frac{k_0(k_0+p_0)+\vec{k}(\vec{k}+\vec{p})-m^2}{(m^2-k^2)(m^2-(k+p)^2)}
\end{align}
As we can see, $\mu_I$ is explicitly included in the coefficient, which provides an intuitive reason for keeping $\mu_I$ in the subtraction part. 
Here, for explaining why subtraction term includes the chemical potential in a more general case, we omit the coefficient $- 4 \mu_I^2 \sin^2\pi $ of Eq.\eqref{eq:loop}.
Since we are taking the mean field approximation, we set the vector component of pion momentum to be zero $\vec{p}=0$ 
and the pion Matsubara frequency to be zero $p_0=\mu_\pi^*=\mu_I^*$. 
Defining the energy $\omega=\sqrt{m^2+{\vec{k}}^2}$ and neglecting the coefficient, this one loop contribution is rewritten as
\begin{equation}
\tilde{\Sigma}=T\sum_{n}\int \frac{d^3k}{(2\pi)^3}\frac{k_0(k_0+\mu_I^*)-\omega^2}{(\omega^2-k_0^2)(\omega^2-(k_0+\mu_I^*)^2)}
\end{equation}
with $k_0=i\omega^F_n+\mu_n^*$.
Following the standard procedure, we take
the summation over the Matsubara frequency as
\begin{align}
&T\sum_{n} f\left(k_{0}=i \omega_{n}^F+\mu_n^*\right) \notag \\ = & -\frac{1}{2 \pi i} \int_{-i \infty+\mu+\epsilon}^{i \infty+\mu+\epsilon} d k_{0} f\left(k_{0}\right) \frac{1}{\mathrm{e}^{\beta\left(k_{0}-\mu\right)}+1} \notag \\
& -\frac{1}{2 \pi i} \int_{-i \infty+\mu-\epsilon}^{i \infty+\mu-\epsilon} d k_{0} f\left(k_{0}\right) \frac{1}{\mathrm{e}^{\beta\left(\mu-k_{0}\right)}+1} \notag \\
& +\frac{1}{2 \pi i} \oint_{C} d k_{0} f\left(k_{0}\right)+\frac{1}{2 \pi i} \int_{-i \infty}^{i \infty} d k_{0} f\left(k_{0}\right),
\label{eq:standardpro}
\end{align}
where
\begin{equation}
f(k_0)=\int \frac{d^3k}{(2\pi)^3} \frac{k_0(k_0+\mu_I^*)-\omega^2}{(\omega^2-k_0^2)(\omega^2-(k_0+\mu_I^*)^2)}.
\label{f0 function}
\end{equation}
We note that, in the standard procedure where only one chemical potential exists, $f\left(k_0\right)$ does not depend on the chemical potential. In the present case, on the other hand, two different chemical potentials exist, so that  $f\left(k_0\right)$ does include the chemical potential as shown in Eq.~(\ref{f0 function}).
The first and second term in Eq.~\ref{eq:standardpro} are the contributions from particles and antiparticles at non-zero temperature.
They will disappear when we take zero-temperature limit. The third term is the contribution from finite density which is the main contribution to the thermodynamic potential. 
%\cmh{[! The above procedure is for the standard case where only one chemical potential is included. The procedure below is for the case where 2 different chemical potentials are included.You should stress that you are considering the case with 2 different chemical potentials. !]}
The last term is the vacuum contribution that we subtract. Here it is
\begin{equation}
\Sigma_{\rm sub}
%\Omega_{\rm sub}
= \frac{1}{2\pi i} \int \frac{d^3k}{(2\pi)^3} \int_{-i\infty}^{i\infty}dk_0\frac{k_0(k_0+\mu_I^*)-\omega^2}{(\omega^2-k_0^2)(\omega^2-(k_0+\mu_I^*)^2)}
\label{eq:sub}
\end{equation}

In the normal phase where pion condensation does not exist and only $\sigma$, $\omega$, and $\rho$ mean fields exist, no loop contribution contains both protons and neutrons; it contains only one or the other. 
Then, we can easily apply the no-sea approximation to subtract the vacuum contribution by simply setting the chemical potential to be zero. However, when the pion condensate exits, as we can see in the Eq.~\ref{eq:sub}, the difference of the two chemical potentials $\mu_I^*=\mu_p^*-\mu_n^*$ is explicitly included in the subtraction term. 
This is the reason why the subtraction term include the isospin chemical potential $\mu_I$ as $E_i(0,\mu_I,\phi)$ in Eq.~(\ref{subtracted Omega}).
We numerically checked that this subtraction actually removes the divergence.
\begin{widetext}

\section{Derivation of effective mass and chemical potential for $\pi$}
\label{appendix:pion}
In this appendix, we derive the effective mass and chemical potential of pion. By using non-linear decomposition given in Eq.(\ref{M decomp}) and Eq.(\ref{eq:non-linear pi}) with $\sum_a\pi_aT^a$ field replaced by $\sum_a \frac{\pi_aT^a}{\sigma}=\frac{\pi}{\sigma}$, the mesonic Lagrangian is rewritten as
\begin{align}
\mathcal{L}_\pi &= \sigma^2 \operatorname{tr}[\hat{\alpha}_{\mu\perp}\hat{\alpha}^{\mu}_{\perp}]+\frac{m_{\omega}^{2}}{2 g_{\omega}^{2}} \operatorname{tr}\left[\hat{\alpha}_{\mu}^{\|}\right] \operatorname{tr}\left[\hat{\alpha}_{\|}^{\mu}\right]+\frac{m_{\rho}^{2}}{g_{\rho}^{2}}\left(\operatorname{tr}\left[\hat{\alpha}_{\mu}^{\|} \hat{\alpha}_{\|}^{\mu}\right]-\frac{1}{2} \operatorname{tr}\left[\hat{\alpha}_{\mu}^{\|}\right] \operatorname{tr}\left[\hat{\alpha}_{\|}^{\mu}\right]\right) \notag \\
&=\sigma^2 \operatorname{tr}\left[\left(\frac{1}{\sigma}\partial_\mu \pi -\frac{i}{\sigma}[\mathcal{V}^\mu,\pi] \right)^2 \right]+\frac{m_\rho^2}{g_\rho^2}\operatorname{tr}\left[ \left(-g_\rho \rho_\mu+\mathcal{V_\mu}+\frac{i}{2\sigma}[\partial_\mu\pi,\pi] \right)^2 \right] + \cdots\notag \\
&=\operatorname{tr}\left[\left(\partial_\mu\pi-i[\mathcal{V}_\mu,\pi]+\frac{m_\rho^2i}{2g_\rho^2 \sigma^2} [\rho_\mu-\frac{1}{g_\rho}\mathcal{V}_\mu, \pi]\right)^2\right] + \cdots
\end{align}
where $\mathcal{V}_\mu=\mu_IT^3\delta\mu_{0\mu}$. From the definition of $\rho$ mean field given in Eq.(\ref{eq:mean fields definition}), the effective pion chemical potential is expressed as
\begin{equation}
\mu_\pi^*=\mu_I-\frac{m_\rho^2}{2g_\rho^2 \sigma^2}\rho.
\end{equation}

The explicit breaking term $V_{SB}$ given in Eq.(\ref{eq:VSB}) is rewritten as
\begin{equation}
V_{SB}={m_\pi^*}^2 \operatorname{tr}[\pi^2]+\cdots,
\end{equation}
where
\begin{equation}
{m_\pi^*}^2=\frac{f_\pi m_\pi^2}{\sigma}.
\end{equation}
\end{widetext}
\bibliography{main}

@article{Minamikawa:2023eky,
    author = "Minamikawa, Takuya and Gao, Bikai and Kojo, Toru and Harada, Masayasu",
    title = "{Chiral Restoration of Nucleons in Neutron Star Matter: Studies Based on a Parity Doublet Model}",
    eprint = "2302.00825",
    archivePrefix = "arXiv",
    primaryClass = "nucl-th",
    doi = "10.3390/sym15030745",
    journal = "Symmetry",
    volume = "15",
    number = "3",
    pages = "745",
    year = "2023"
}

@article{Minamikawa:2020jfj,
    author = "Minamikawa, Takuya and Kojo, Toru and Harada, Masayasu",
    title = "{Quark-hadron crossover equations of state for neutron stars: constraining the chiral invariant mass in a parity doublet model}",
    eprint = "2011.13684",
    archivePrefix = "arXiv",
    primaryClass = "nucl-th",
    doi = "10.1103/PhysRevC.103.045205",
    journal = "Phys. Rev. C",
    volume = "103",
    number = "4",
    pages = "045205",
    year = "2021"
}

@article{Gao:2022klm,
    author = "Gao, Bikai and Minamikawa, Takuya and Kojo, Toru and Harada, Masayasu",
    title = "{Impacts of the U(1)A anomaly on nuclear and neutron star equation~of state based on a parity doublet model}",
    eprint = "2207.05970",
    archivePrefix = "arXiv",
    primaryClass = "nucl-th",
    doi = "10.1103/PhysRevC.106.065205",
    journal = "Phys. Rev. C",
    volume = "106",
    number = "6",
    pages = "065205",
    year = "2022"
}

@article{Detar:1988kn,
    author = "Detar, Carleton E. and Kunihiro, Teiji",
    title = "{Linear $\sigma$ Model With Parity Doubling}",
    reportNumber = "RIFP-748",
    doi = "10.1103/PhysRevD.39.2805",
    journal = "Phys. Rev. D",
    volume = "39",
    pages = "2805",
    year = "1989"
}

@article{Aarts:2015mma,
    author = {Aarts, Gert and Allton, Chris and Hands, Simon and J\"ager, Benjamin and Praki, Chrisanthi and Skullerud, Jon-Ivar},
    title = "{Nucleons and parity doubling across the deconfinement transition}",
    eprint = "1502.03603",
    archivePrefix = "arXiv",
    primaryClass = "hep-lat",
    doi = "10.1103/PhysRevD.92.014503",
    journal = "Phys. Rev. D",
    volume = "92",
    number = "1",
    pages = "014503",
    year = "2015"
}

@article{Aarts:2017rrl,
    author = {Aarts, Gert and Allton, Chris and De Boni, Davide and Hands, Simon and J\"ager, Benjamin and Praki, Chrisanthi and Skullerud, Jon-Ivar},
    title = "{Light baryons below and above the deconfinement transition: medium effects and parity doubling}",
    eprint = "1703.09246",
    archivePrefix = "arXiv",
    primaryClass = "hep-lat",
    doi = "10.1007/JHEP06(2017)034",
    journal = "JHEP",
    volume = "06",
    pages = "034",
    year = "2017"
}

@article{Aarts:2018glk,
    author = {Aarts, Gert and Allton, Chris and De Boni, Davide and J\"ager, Benjamin},
    title = "{Hyperons in thermal QCD: A lattice view}",
    eprint = "1812.07393",
    archivePrefix = "arXiv",
    primaryClass = "hep-lat",
    reportNumber = "CP3-Origins-2018-050 DNRF90",
    doi = "10.1103/PhysRevD.99.074503",
    journal = "Phys. Rev. D",
    volume = "99",
    number = "7",
    pages = "074503",
    year = "2019"
}

@article{Jido:2001nt,
    author = "Jido, Daisuke and Oka, Makoto and Hosaka, Atsushi",
    title = "{Chiral symmetry of baryons}",
    eprint = "hep-ph/0110005",
    archivePrefix = "arXiv",
    doi = "10.1143/PTP.106.873",
    journal = "Prog. Theor. Phys.",
    volume = "106",
    pages = "873--908",
    year = "2001"
}

@article{Motohiro:2015taa,
    author = "Motohiro, Yuichi and Kim, Youngman and Harada, Masayasu",
    title = "{Asymmetric nuclear matter in a parity doublet model with hidden local symmetry}",
    eprint = "1505.00988",
    archivePrefix = "arXiv",
    primaryClass = "nucl-th",
    doi = "10.1103/PhysRevC.92.025201",
    journal = "Phys. Rev. C",
    volume = "92",
    number = "2",
    pages = "025201",
    year = "2015",
    note = "[Erratum: Phys.Rev.C 95, 059903 (2017)]"
}

@article{Yamazaki:2019tuo,
    author = "Yamazaki, Takahiro and Harada, Masayasu",
    title = "{Constraint to chiral invariant masses of nucleons from GW170817 in an extended parity doublet model}",
    eprint = "1901.02167",
    archivePrefix = "arXiv",
    primaryClass = "nucl-th",
    doi = "10.1103/PhysRevC.100.025205",
    journal = "Phys. Rev. C",
    volume = "100",
    number = "2",
    pages = "025205",
    year = "2019"
}

@article{Li:2021thg,
    author = "Li, Bao-An and Cai, Bao-Jun and Xie, Wen-Jie and Zhang, Nai-Bo",
    title = "{Progress in Constraining Nuclear Symmetry Energy Using Neutron Star Observables Since GW170817}",
    eprint = "2105.04629",
    archivePrefix = "arXiv",
    primaryClass = "nucl-th",
    doi = "10.3390/universe7060182",
    journal = "Universe",
    volume = "7",
    number = "6",
    pages = "182",
    year = "2021"
}

@article{BANDO1988217,
title = {Nonlinear realization and hidden local symmetries},
journal = {Physics Reports},
volume = {164},
number = {4},
pages = {217-314},
year = {1988},
issn = {0370-1573},
doi = {https://doi.org/10.1016/0370-1573(88)90019-1},
url = {https://www.sciencedirect.com/science/article/pii/0370157388900191},
author = {Masako Bando and Taichiro Kugo and Koichi Yamawaki}
}

@article{HARADA20031,
title = {Hidden local symmetry at loop: A new perspective of composite gauge boson and chiral phase transition},
journal = {Physics Reports},
volume = {381},
number = {1},
pages = {1-233},
year = {2003},
issn = {0370-1573},
doi = {https://doi.org/10.1016/S0370-1573(03)00139-X},
url = {https://www.sciencedirect.com/science/article/pii/S037015730300139X},
author = {Masayasu Harada and Koichi Yamawaki}
}

@article{Minamikawa:2021fln,
    author = "Minamikawa, Takuya and Kojo, Toru and Harada, Masayasu",
    title = "{Chiral condensates for neutron stars in hadron-quark crossover: From a parity doublet nucleon model to a Nambu\textendash{}Jona-Lasinio quark model}",
    eprint = "2107.14545",
    archivePrefix = "arXiv",
    primaryClass = "nucl-th",
    doi = "10.1103/PhysRevC.104.065201",
    journal = "Phys. Rev. C",
    volume = "104",
    number = "6",
    pages = "065201",
    year = "2021"
}

@article{Minamikawa:2023ypn,
    journal = {},
    author = "Minamikawa, Takuya and Gao, Bikai and Kojo, Toru and Harada, Masayasu",
    title = "{Parity doublet model for baryon octets: diquark classifications and mass hierarchy based on the quark-line diagram}",
    eprint = "2306.15564",
    archivePrefix = "arXiv",
    primaryClass = "hep-ph",
    month = "6",
    year = "2023"
}

@article{Kong:2023nue,
    author = "Kong, Yuk Kei and Minamikawa, Takuya and Harada, Masayasu",
    title = "{Neutron star matter based on a parity doublet model including the a0(980) meson}",
    eprint = "2306.08140",
    archivePrefix = "arXiv",
    primaryClass = "nucl-th",
    doi = "10.1103/PhysRevC.108.055206",
    journal = "Phys. Rev. C",
    volume = "108",
    number = "5",
    pages = "055206",
    year = "2023"
}

@article{Son:2000xc,
    author = "Son, D. T. and Stephanov, Misha A.",
    title = "{QCD at finite isospin density}",
    eprint = "hep-ph/0005225",
    archivePrefix = "arXiv",
    doi = "10.1103/PhysRevLett.86.592",
    journal = "Phys. Rev. Lett.",
    volume = "86",
    pages = "592--595",
    year = "2001"
}

@article{PhysRevD.103.L051501,
  title = {Vector meson mass in the chiral symmetry restored vacuum},
  author = {Kim, Jisu and Lee, Su Houng},
  journal = {Phys. Rev. D},
  volume = {103},
  issue = {5},
  pages = {L051501},
  numpages = {5},
  year = {2021},
  month = {Mar},
  publisher = {American Physical Society},
  doi = {10.1103/PhysRevD.103.L051501},
  url = {https://link.aps.org/doi/10.1103/PhysRevD.103.L051501}
}

@article{PhysRevD.105.014014,
  title = {Masses of hadrons in the chiral symmetry restored vacuum},
  author = {Kim, Jisu and Lee, Su Houng},
  journal = {Phys. Rev. D},
  volume = {105},
  issue = {1},
  pages = {014014},
  numpages = {15},
  year = {2022},
  month = {Jan},
  publisher = {American Physical Society},
  doi = {10.1103/PhysRevD.105.014014},
  url = {https://link.aps.org/doi/10.1103/PhysRevD.105.014014}
}

@Article{sym15040799,
AUTHOR = {Lee, Su Houng},
TITLE = {Chiral Symmetry Breaking and the Masses of Hadrons: A Review},
JOURNAL = {Symmetry},
VOLUME = {15},
YEAR = {2023},
NUMBER = {4},
ARTICLE-NUMBER = {799},
URL = {https://www.mdpi.com/2073-8994/15/4/799},
ISSN = {2073-8994},
DOI = {10.3390/sym15040799}
}

@article{Kojo:2024sca,
    author = "Kojo, Toru and Suenaga, Daiki and Chiba, Ryuji",
    title = "{Isospin QCD as a Laboratory for Dense QCD}",
    eprint = "2406.11059",
    archivePrefix = "arXiv",
    primaryClass = "hep-ph",
    doi = "10.3390/universe10070293",
    journal = "Universe",
    volume = "10",
    number = "7",
    pages = "293",
    year = "2024"
}

@article{94zt-q2dw,
  title = {Pion condensation and pion star from holographic QCD},
  author = {Chen, Yidian and Ding, Mingshan and Li, Danning and Fadafan, Kazem Bitaghsir and Huang, Mei},
  journal = {Phys. Rev. D},
  volume = {111},
  issue = {12},
  pages = {126010},
  numpages = {13},
  year = {2025},
  month = {Jun},
  publisher = {American Physical Society},
  doi = {10.1103/94zt-q2dw},
  url = {https://link.aps.org/doi/10.1103/94zt-q2dw}
}

@article{Voskresensky:2022gts,
    author = "Voskresensky, D. N.",
    title = "{S-wave pion condensation in symmetric nuclear matter}",
    eprint = "2201.07536",
    archivePrefix = "arXiv",
    primaryClass = "nucl-th",
    doi = "10.1103/PhysRevD.105.116007",
    journal = "Phys. Rev. D",
    volume = "105",
    number = "11",
    pages = "116007",
    year = "2022"
}

@article{Fore:2023gwv,
    author = "Fore, Bryce and Kaiser, Norbert and Reddy, Sanjay and Warrington, Neill C.",
    title = "{Mass of charged pions in neutron-star matter}",
    eprint = "2301.07226",
    archivePrefix = "arXiv",
    primaryClass = "nucl-th",
    doi = "10.1103/PhysRevC.110.025803",
    journal = "Phys. Rev. C",
    volume = "110",
    number = "2",
    pages = "025803",
    year = "2024"
}

@article{Muto:2024upf,
    author = "Muto, Takumi",
    title = "{Properties of a kaon-condensed phase in hyperon-mixed matter with three-baryon forces}",
    eprint = "2411.09967",
    archivePrefix = "arXiv",
    primaryClass = "nucl-th",
    doi = "10.1103/PhysRevC.111.045802",
    journal = "Phys. Rev. C",
    volume = "111",
    number = "4",
    pages = "045802",
    year = "2025"
}

@article{Bhat:2026hco,
    journal = {},
    author = "Bhat, Bhavnesh and Dohi, Akira and Muto, Takumi and Noda, Tsuneo",
    title = "{Cooling of Isolated Neutron Stars with Hyperon-mixed Kaon-Condensation Matter}",
    eprint = "2605.09723",
    archivePrefix = "arXiv",
    primaryClass = "astro-ph.HE",
    reportNumber = "RIKEN-iTHEMS-Report-26",
    month = "5",
    year = "2026"
}

@article{Gao:2025nkg,
    author = "Gao, Bikai and Liu, Xiang and Harada, Masayasu and Ma, Yong-Liang",
    title = "{Implication of neutron star observations to the origin of nucleon mass}",
    eprint = "2508.00243",
    archivePrefix = "arXiv",
    primaryClass = "nucl-th",
    doi = "10.1007/s11433-025-2839-7",
    journal = "Sci. China Phys. Mech. Astron.",
    volume = "69",
    number = "3",
    pages = "232011",
    year = "2026"
}

@article{Gao:2024chh,
    author = "Gao, Bikai and Yan, Yan and Harada, Masayasu",
    title = "{Reconciling constraints from the supernova remnant HESS J1731-347 with the parity doublet model}",
    eprint = "2404.04786",
    archivePrefix = "arXiv",
    primaryClass = "nucl-th",
    doi = "10.1103/PhysRevC.109.065807",
    journal = "Phys. Rev. C",
    volume = "109",
    number = "6",
    pages = "065807",
    year = "2024"
}

@article{Gao:2024mew,
    author = "Gao, Bikai and Kojo, Toru and Harada, Masayasu",
    title = "{Parity doublet model for baryon octets: Ground states saturated by good diquarks and the role of bad diquarks for excited states}",
    eprint = "2403.18214",
    archivePrefix = "arXiv",
    primaryClass = "hep-ph",
    doi = "10.1103/PhysRevD.110.016016",
    journal = "Phys. Rev. D",
    volume = "110",
    number = "1",
    pages = "016016",
    year = "2024"
}

@article{Gao:2026jsg,
    journal = {},
    author = "Gao, Bikai and Harada, Masayasu",
    title = "{A quark-diquark model for parity doublet structure of baryons}",
    eprint = "2607.25178",
    archivePrefix = "arXiv",
    primaryClass = "hep-ph",
    month = "7",
    year = "2026"
}

@article{Gao:2026scv,
    author = "Gao, Bikai",
    title = "{Chiral symmetry restoration and hyperon suppression in neutron stars}",
    eprint = "2602.12503",
    archivePrefix = "arXiv",
    primaryClass = "nucl-th",
    doi = "10.1103/vb5r-vdm7",
    journal = "Phys. Rev. D",
    volume = "113",
    number = "8",
    pages = "083012",
    year = "2026"
}

@article{Brodie:2025nww,
    author = "Brodie, Liam and Pisarski, Robert D.",
    title = "{Parity-Doubled Nucleons Can Rapidly Cool Neutron Stars}",
    eprint = "2501.02055",
    archivePrefix = "arXiv",
    primaryClass = "nucl-th",
    doi = "10.1103/bf66-1hr5",
    journal = "Phys. Rev. Lett.",
    volume = "135",
    number = "15",
    pages = "152702",
    year = "2025"
}

@article{Negreiros:2026ode,
    journal = {},
    author = "Negreiros, Rodrigo and Brodie, Liam and Steinheimer, Jan and Dexheimer, Veronica and Pisarski, Robert D.",
    title = "{Enhanced Neutrino Cooling from Parity-Doubled Nucleons in Neutron Star Cooling Simulations}",
    eprint = "2603.06789",
    archivePrefix = "arXiv",
    primaryClass = "astro-ph.HE",
    month = "3",
    year = "2026"
}

@article{Nishihara:2014nva,
    author = "Nishihara, Hiroki and Harada, Masayasu",
    title = "{Enhancement of Chiral Symmetry Breaking from the Pion condensation at finite isospin chemical potential in a holographic QCD model}",
    eprint = "1401.2928",
    archivePrefix = "arXiv",
    primaryClass = "hep-ph",
    doi = "10.1103/PhysRevD.89.076001",
    journal = "Phys. Rev. D",
    volume = "89",
    number = "7",
    pages = "076001",
    year = "2014"
}

@article{Sawyer:1972cq,
    author = "Sawyer, R. F.",
    title = "{Condensed pi- phase in neutron star matter}",
    doi = "10.1103/PhysRevLett.29.382",
    journal = "Phys. Rev. Lett.",
    volume = "29",
    pages = "382--385",
    year = "1972"
}

@article{Scalapino:1972fu,
    author = "Scalapino, D. J.",
    title = "{Pi- condensate in dense nuclear matter}",
    doi = "10.1103/PhysRevLett.29.386",
    journal = "Phys. Rev. Lett.",
    volume = "29",
    pages = "386--388",
    year = "1972"
}

@article{Migdal:1973zm,
    author = "Migdal, A. B.",
    title = "{Pi condensation in nuclear matter}",
    doi = "10.1103/PhysRevLett.31.257",
    journal = "Phys. Rev. Lett.",
    volume = "31",
    pages = "257--260",
    year = "1973"
}

@article{Baym:1974vzp,
    author = "Baym, Gordon and Flowers, Elliott",
    title = "{Pion condensation in neutron star matter: Equilibrium conditions and model calculations}",
    doi = "10.1016/0375-9474(74)90583-1",
    journal = "Nucl. Phys. A",
    volume = "222",
    pages = "29--64",
    year = "1974"
}

@article{BROWN19761,
title = {Pion condensates},
journal = {Physics Reports},
volume = {27},
number = {1},
pages = {1-34},
year = {1976},
issn = {0370-1573},
doi = {https://doi.org/10.1016/0370-1573(76)90008-9},
url = {https://www.sciencedirect.com/science/article/pii/0370157376900089},
author = {G.E. Brown and W. Weise},
}

@article{Motta:2026pby,
    author = "Motta, Theo F. and Pradinett, Randall H. V. and Krein, Gast{\~a}o",
    title = "{Hyperon-induced inhomogeneous pion condensation and moat regimes in neutron star cores}",
    eprint = "2602.23433",
    archivePrefix = "arXiv",
    primaryClass = "nucl-th",
    doi = "10.1103/4cv9-wlkn",
    journal = "Phys. Rev. D",
    volume = "113",
    number = "9",
    pages = "094028",
    year = "2026"
}

\end{document}